\documentclass[fleqn,usenatbib]{mnras}

\usepackage{newtxtext,newtxmath}

\usepackage[T1]{fontenc}

\DeclareRobustCommand{\VAN}[3]{#2}
\let\VANthebibliography\thebibliography
\def\thebibliography{\DeclareRobustCommand{\VAN}[3]{##3}\VANthebibliography}

\usepackage{graphicx}	
\usepackage{amsmath}	

\usepackage{scalerel,tikz}
\usetikzlibrary{svg.path}
\definecolor{orcidlogocol}{HTML}{A6CE39}
\tikzset{orcidlogo/.pic={
 \fill[orcidlogocol] svg{M256,128c0,70.7-57.3,128-128,128C57.3,256,0,198.7,0,128C0,57.3,57.3,0,128,0C198.7,0,256,57.3,256,128z};
 \fill[white] svg{M86.3,186.2H70.9V79.1h15.4v48.4V186.2z}
 svg{M108.9,79.1h41.6c39.6,0,57,28.3,57,53.6c0,27.5-21.5,53.6-56.8,53.6h-41.8V79.1z M124.3,172.4h24.5c34.9,0,42.9-26.5,42.9-39.7c0-21.5-13.7-39.7-43.7-39.7h-23.7V172.4z}
 svg{M88.7,56.8c0,5.5-4.5,10.1-10.1,10.1c-5.6,0-10.1-4.6-10.1-10.1c0-5.6,4.5-10.1,10.1-10.1C84.2,46.7,88.7,51.3,88.7,56.8z};
}}
\newcommand\orcidicon[1]{\href{https://orcid.org/#1}{\mbox{\scalerel*{
\begin{tikzpicture}[yscale=-1,transform shape]
\pic{orcidlogo};
\end{tikzpicture}
}{|}}}}

\newcommand{\mr}{\mathrm}
\newcommand{\als}{\alpha_{\mathrm{vir}}}
\usepackage{multirow}

\title[Influence of cloud virial parameter]{The influence of the cloud virial parameter on the initial mass function}

\author[Mathew, Federrath, \& Seta]{
Sajay Sunny Mathew$^{\orcidicon{0000-0002-8381-8195}\,1}$\thanks{E-mail: \href{mailto:sajay.mathew@anu.edu.au}{sajay.mathew@anu.edu.au}},
Christoph Federrath$^{\orcidicon{0000-0002-0706-2306}\,1,2}$\thanks{E-mail: \href{mailto:christoph.federrath@anu.edu.au}{christoph.federrath@anu.edu.au}}, and
Amit Seta$^{\orcidicon{0000-0001-9708-0286}\,1}$\thanks{E-mail: \href{mailto:amit.seta@anu.edu.au}{amit.seta@anu.edu.au}}
\\
$^{1}$Research School of Astronomy and Astrophysics, Australian National University, Canberra, ACT~2611, Australia\\
$^{2}$Australian Research Council Centre of Excellence in All Sky Astrophysics (ASTRO3D), Canberra, ACT~2611, Australia
}

\date{Accepted XXX. Received YYY; in original form ZZZ}

\pubyear{\the\year{}}

\begin{document}
\label{firstpage}
\pagerange{\pageref{firstpage}--\pageref{lastpage}}
\maketitle

\begin{abstract}
Crucial for star formation is the interplay between gravity and turbulence. The observed cloud virial parameter, $\alpha_{\mathrm{vir}}$, which is the ratio of twice the turbulent kinetic energy to the gravitational energy, is found to vary significantly in different environments, where the scatter among individual star-forming clouds can exceed an order of magnitude. Therefore, a strong dependence of the initial mass function (IMF) on $\alpha_{\mathrm{vir}}$ may challenge the notion of a universal IMF. To determine the role of $\alpha_{\mathrm{vir}}$ on the IMF, we compare the star-particle mass functions obtained in high-resolution magnetohydrodynamical simulations including jet and heating feedback, with $\alpha_{\mathrm{vir}}=0.0625$, $0.125$, and $0.5$. We find that varying $\als$ from $\alpha_{\mathrm{vir}}\sim0.5$ to $\alpha_{\mathrm{vir}}<0.1$ shifts the peak of the IMF to lower masses by a factor of $\sim2$ and increases the star formation rate by a similar factor. The dependence of the IMF and star formation rate on $\alpha_{\mathrm{vir}}$ is non-linear, with the dependence subsiding at $\alpha_{\mathrm{vir}}<0.1$. Our study shows a systematic dependence of the IMF on $\alpha_{\mathrm{vir}}$. Yet, it may not be measurable easily in observations, considering the uncertainties, and the relatively weak dependence found in this study.  
\end{abstract}




\section{Introduction}
One of the fundamental questions in star formation is the initial mass function (IMF), which is the stellar mass distribution in a newly formed cluster. The IMF is considered nearly universal based on observations \citep[see reviews by][]{2010ARA&A..48..339B,2013pss5.book..115K,2014PhR...539...49K,2014prpl.conf...53O,2018PASA...35...39H,2020SSRv..216...70L,2024arXiv240407301H}. Despite some variations in the functional form, most observational IMF models identify the existence of a peak at around $0.3\, \mr{M_\odot}$ and a power-law tail at high masses given by $dN \propto M^{-\Gamma}\, d\mathrm{log}M$, where $N$ is the number of stars and $M$ is the stellar mass \citep{1955ApJ...121..161S,1979ApJS...41..513M,2001MNRAS.322..231K,chabrier2005,2011ApJ...726...27P}. The slope (power-law on a logarithmic scale) of the high-mass end measured in the Milky Way and nearby galaxies is generally consistent with the \citet{1955ApJ...121..161S} slope \citep{2002Sci...295...82K,2010ARA&A..48..339B,2018PASA...35...39H,2013pss5.book..115K}, although cluster-to-cluster variations exist \citep{2014MNRAS.444.1957D,2017MNRAS.464.1738D,2018A&A...618A..73S,2024arXiv240407301H,2024arXiv241007311K}. 

Recent observations challenge the notion of a universal IMF. There is compelling evidence for a top-heavy IMF near the Galactic center, i.e, having a higher fraction of high-mass stars compared to the IMF in the solar neighbourhood \citep[e.g.,][]{1999ApJ...525..750F,2006ApJ...653L.113K,2013ApJ...764..155L,2019ApJ...870...44H}. On the other hand, massive elliptical early-type galaxies (ETGs) are discovered to have a bottom-heavy IMF, i.e, a higher fraction of low-mass stars \citep[e.g.,][]{2003MNRAS.339L..12C,2010Natur.468..940V,2010ApJ...709.1195T,2012Natur.484..485C,2013ApJ...776L..26C}. Further, the slope of the IMF seems to be dependent on the star formation rate (SFR) of the galaxy, where the power-law slope becomes shallower with the increase in SFR \citep{2011MNRAS.415.1647G}. In addition, \citet{2013pss5.book..115K} suggest that the underlying formation mechanism of brown dwarfs (BDs) and stars are likely different based on the observed BD binary statistics and propose separate mass functions for BDs and stellar objects. The need to treat the mass functions for BDs and stars separately would put further constraints on the modelling of the IMF parameters and interpreting the IMF shape in the sub-solar range. An accurate theoretical model of the IMF must reproduce the typical IMF characteristics such as the presence of a peak in the low-mass end, the existence of a power-law tail at super-solar masses with a slope close to the \citet{1955ApJ...121..161S} estimate, and at the same time address the observed outlier IMFs, apparent SFR dependence, and the BD-star issue discussed above. A prerequisite to achieving such a model is understanding the dependence of the IMF on environmental conditions.

There have been significant contributions from theoretical and numerical works towards understanding the IMF. Some of the proposed candidates responsible for setting the peak mass include the Jeans mass at the density of the thermal coupling of gas to the dust \citep{2003ApJ...592..975L,2005MNRAS.359..211L,2005A&A...435..611J,2008ApJ...681..365E,2023arXiv230816268G}, the Jeans mass associated with protostellar heating \citep{bate2009,2011ApJ...740...74K,2016MNRAS.458..673G,2017JPhCS.837a2007F,mathew2020,hennebelle2020}, and the formation of the first hydrostatic core and tidal screening of the gas in the immediate environment by the core \citep{2019ApJ...883..140H,2020MNRAS.492.4727C}. Another important mechanism extensively studied in the context of the IMF is the role of magnetic fields, which is found to reduce fragmentation by providing additional pressure support \citep{price2008magnetic,2015MNRAS.450.4035F}, although the extent of influence is debatable in the presence of protostellar feedback \citep{2014MNRAS.439.3420M,2019FrASS...6....7K}. Protostellar jets/outflows, which are driven by magnetic fields, induce fragmentation and reduce the median mass by a factor of $\sim 2$--3 \citep{2006ApJ...640L.187L,2010ApJ...709...27W,2014ApJ...790..128F,2014prpl.conf..451F,2017ApJ...847..104O,mathew2021,2021MNRAS.502.3646G,2024A&A...683A..13L}. Further, the impact of magnetic fields may differ for super-Alfv$\acute{\text{e}}$nic and trans-Alfv$\acute{\text{e}}$nic conditions, and also in a multi-phase medium \citep{1998PhRvL..80.2754M,2019FrASS...6....7K,2021MNRAS.504.4354B,2022MNRAS.514..957S}. Assuming a correlation between the core mass function (CMF) and the IMF, supported by some observations \citep{1998A&A...336..150M,1998ApJ...508L..91T,2000ApJ...545..327J,2006A&A...447..609S,2007A&A...462L..17A,2015A&A...584A..91K}, analytical theories of the CMF/IMF \citep{2002ApJ...576..870P,2008ApJ...684..395H,2012MNRAS.423.2037H} based on gravo-turbulent fragmentation suggest that the IMF is sensitive to the turbulent properties of the cloud, such as the Mach number and the mode of turbulence driving. Meanwhile, star formation and IMF theories based on a global hierarchical collapse \citep[e.g.,][]{2007prpl.conf...63B,2017MNRAS.467.1313V,2019MNRAS.490.3061V} and stochastic, competitive accretion are also prominent \citep[e.g.,][]{1982NYASA.395..226Z,2001MNRAS.324..573B,bateandbonnell2005,2015MNRAS.449.2413B}. Recently, there have been numerous studies on the IMF in the early universe, with a focus on the influence of varying metallicities \citep[e.g.,][]{2022MNRAS.509.1959S,2023MNRAS.519..688B,2023Natur.613..460L,2024ApJ...969...95Y,2024MNRAS.527.7306T,2024ApJ...967L..28M}.

While the above findings put crucial constraints, we are yet to reach a clear consensus on the origin of the IMF. A primary reason why it is difficult to isolate the mechanism(s) responsible for the observed IMF characteristics is because the effects of the mechanisms involved in star formation are inter-related. Thus, it is important to include all these mechanisms simultaneously and analyse the impact of the relative parameters associated with the cloud. The current analytical theories of the CMF/IMF are fundamentally based on the interaction between gravity and turbulence \citep{2004RvMP...76..125M}. The principal quantity that defines the relative importance of gravity vs.~turbulence is the cloud virial parameter $\als$, defined as twice the ratio of turbulent kinetic to gravitational energy. It is typically found to be around unity, but the scatter in the measurements is large and the difference in $\als$ between individual clouds can be more than an order of magnitude \citep{2013ApJ...779..185K,2024arXiv240711125P}. A strong dependence of the IMF on $\als$ would place the concept of IMF universality under scrutiny. Therefore, we carry out a comprehensive study of the influence of the virial parameter on the star formation process, in particular the IMF, using simulations of star cluster formation. The simulations include gravity, turbulence, magnetic fields, protostellar heating and mechanical feedback in the form of jets and outflows. 

In Section~\ref{sec:methods}, we outline the numerical methodology and the initial conditions of the simulations. In Section~\ref{sec:results}, we study the impact of the cloud virial parameter on the SFR and the IMF. We also compare the IMF and the multiplicity associated with simulations of different virial parameters with observations. In Section~\ref{sec:discussion}, we discuss some of the previous numerical studies on the effect of the cloud virial parameter on the IMF. The main conclusions are presented in Section~\ref{sec:conclusions}.

\section{Methodology}
\label{sec:methods}
\subsection{Numerical setup}
Here we provide a brief summary of the numerical methods used in this study, which build upon the simulation framework in \citet{mathew2021}. A more detailed description of the general setup can be found in \S2 of \citet{mathew2021}. We model star cluster formation by solving the magnetohydrodynamical (MHD) equations in the presence of gravity utilising the \textsc{flash} (version~4) code \citep{2000ApJS..131..273F,2008ASPC..385..145D} with in-house modifications \citep{2011JCoPh.230.3331W}. In addition to gravity and magnetic fields, which are taken into account through the MHD equations, we also implement other important mechanisms such as turbulence, protostellar heating and jets/outflows in our simulations. 

\subsubsection{Turbulence driving}
We employ a stochastic Ornstein-Uhlenbeck (OU) process \citep{1988CF.....16..257E} to drive turbulence in the simulations, which generates an acceleration field that is included as a source term in the momentum equation of MHD \citep{2010A&A...512A..81F,2022ascl.soft04001F}. The turbulent energy is injected only on the largest scales, but it cascades down to smaller scales, producing a velocity power spectrum $\sim k^{-2}$ or equivalently a velocity dispersion -- size relation of $\sigma_v \propto \ell^{1/2}$ \citep{1981MNRAS.194..809L,2002A&A...390..307O,2004ApJ...615L..45H,2011ApJ...740..120R,2013MNRAS.436.1245F,2021NatAs...5..365F}. In this study, we use a natural mixture of turbulence driving modes, i.e., equal power in compressive and solenoidal modes, which corresponds to a turbulence driving parameter of $b \sim 0.4$ \citep[see][for details on the turbulence driving method adopted here]{2008ApJ...688L..79F,2010A&A...512A..81F,2022ascl.soft04001F}.

\subsubsection{Sink particles and protostellar feedback}
We use sink particles to replace the gas within the innermost parts of the collapsing regions in the simulations, which is a common practice in numerical works \citep{1995MNRAS.277..362B,2004ApJ...611..399K,2010ApJ...713..269F}. The introduction of sink particles is to prevent the simulations from reaching extremely small time steps and stalling as a consequence of the rapid increase in the density within the bound regions. We set the sink particle radius $r_{\mathrm{sink}} = 250\, \mr{AU}$, i.e., sink particles will replace spherical volumes of gas with radius $r_{\mathrm{sink}} = 250\, \mr{AU}$ that satisfy several gravitational collapse and sink particle formation criteria \citep{2010ApJ...713..269F}, to avoid spurious formation of star particles. The sink particles inherit the position, linear momentum, and angular momentum of the gas enclosed within the respective spherical volume. The sink particles in our simulations therefore represent star+disc systems. To model protostellar heating and outflow feedback, we use sub-resolution models, which are approximations based on previous theoretical, numerical and observational studies. We encourage the reader to refer to \citet{2014ApJ...790..128F} and \citet{mathew2020,mathew2021} for the full description of the outflow and heating feedback models, respectively. 

\subsubsection{Limitations}
We note that the minimum attainable grid cell size in our simulations is $100\,\mr{AU}$ (see \S\ref{sec: initial conditions}) and the sink particles correspond to star-disc systems as mentioned above. While fragmentation in the extended discs \citep{2002MNRAS.332L..65B,2007A&A...466..943G,2007MNRAS.382L..30S,2011ApJ...730...32S,2012MNRAS.423.1896R,2015ApJ...800...72T} can occur in our simulations, fragmentation on typical disc scales \citep[see reviews by][and the references therein]{2016ARA&A..54..271K,2020SSRv..216...70L} is not fully resolved. However, we aim to analyse the relative differences in the IMFs on changing the virial parameter, which does not require resolving fragmentation entirely on all scales. Moreover, our comparison analysis with observations is aimed at the system IMFs (unresolved close binaries) rather than the canonical (individual star IMF).

\subsection{Initial conditions}
\label{sec: initial conditions}
The computational domain of the simulations is a three-dimensional box with side length $L = 2\, \mr{pc}$ and periodic boundary conditions. The grid structure embodies an adaptive mesh refinement (AMR) framework with a maximum effective resolution of $4096^3$ cells or equivalently a minimum cell size of $100\,$AU. The initial gas density $\rho_0$ and the magnetic field strength $B_0$ in the simulations are uniform. The magnetic field is directed only along the $z$-axis initially. The initial cloud temperature is isothermal at $10\, \mr{K}$, but we employ a polytropic equation of state (EOS) such that the gas pressure (and the temperature via the ideal gas equation) is dependent on the local density \citep[see][for a detailed description of the polytropic EOS implemented here]{mathew2021}. The polytropic EOS is based on previous radiation-hydrodynamical simulations and theoretical works \citep{1969MNRAS.145..271L,1993ApJ...411..274Y,2000ApJ...531..350M,2009ApJ...703..131O}. The steady state sonic Mach number is set as $\mathcal{M} = \sigma_{\mr{v}}/c_{\mr{s}} = 5$, where $\sigma_{\mr{v}} = 1.0\, \mr{km\, s^{-1}}$ is the velocity dispersion on the driving scale and $c_{\mr{s}} = 0.2\, \mr{km\, s^{-1}}$ is the isothermal sound speed. Once turbulence driving starts, filamentary and clump-like over-dense regions start to emerge via turbulent shocks. The stirring of the gas also makes the magnetic field morphology inhomogeneous due to the tangling, compression, and elongation of magnetic field lines by the turbulent flow \citep{2020PhRvF...5d3702S,2021PhRvF...6j3701S}, producing a field structure comparable to that in actual molecular clouds \citep{2016JPlPh..82f5301F}.

To allow turbulence to fully develop and reach a steady state, we drive turbulence in the simulations without self-gravity for two turbulent crossing times, $2\,t_\mathrm{turb}=L/(\mathcal{M}c_\mathrm{s}) = 2\,\mathrm{Myr}$ \citep{2010A&A...512A..81F,pricefederrath2010}. Once a steady state of turbulence is reached, we turn on self-gravity, which we define as time $t = 0$. We allow the simulations to progress until a star formation efficiency (SFE) of 5\% is reached, i.e., when 5\% of the total cloud mass has been converted into stars, during which the turbulence driving is also sustained \citep{1999ApJ...524..169M}. Throughout this phase of evolution, the over-dense regions (analogous to dense cores) in the shocked gas become unstable and form stellar clusters.

With the numerical setup and initial conditions discussed above, we carry out three sets of simulations with different initial cloud virial parameters. For a homogeneous spherical cloud of radius $R_{\mr{cl}}$, total mass $M$, and 3D turbulent velocity dispersion $\sigma_v$, the cloud virial parameter is given by
\begin{equation}
    \als = \frac{2E_{\mathrm{turb}}}{|E_{\mathrm{grav}}|} =\frac{5\sigma_v^2R_{\mr{cl}}}{3GM} =
    \frac{5\mathcal{M}^2c_{\mr{s}}^2}{6G\rho_0 L^2},
    \label{eq: alpha_vir}
\end{equation}
where $E_{\mathrm{turb}}=M\sigma_v^2/2$ and $E_{\mathrm{grav}}=-3GM^2/(5R_{\mathrm{cl}})$, with  $R_{\mr{cl}} = L/2$ and $M = \rho_0 L^3$. It is important to highlight that the cloud structure becomes highly inhomogeneous and the actual value of $\als$ at the time when a steady state of turbulence is reached can be significantly different from the simple spherical approximation given by Eq.~(\ref{eq: alpha_vir}) of an a homogeneous, gravitationally isolated cloud \citep{sfr_fk2012}. However, we refer to $\als$ from Eq.~(\ref{eq: alpha_vir}) to simply denote the distinction between the three simulation models.

To investigate the impact of the virial parameter on the cloud evolution, we conduct three sets of simulations with $\als = 0.0625, 0.125$ and $0.5$ by varying the initial mean gas density $\rho_0$ in the simulations (keeping the Mach number $\mathcal{M}$ constant). We adjust the magnetic field strength $B_0$ in each of the three simulation sets to keep the Alfv\'en Mach number $\mathcal{M_\mr{A}}$ fixed. Maintaining constant values of $\mathcal{M_\mr{A}}$ across the three models is crucial to ensure that the observed variations are independent of the magnetic fields \citep{1998PhRvL..80.2754M,2019FrASS...6....7K,2021MNRAS.504.4354B}. We emphasise that rather than the absolute quantities like $\sigma_v$ and $B_0$, the relative (dimensionless) quantities like $\mathcal{M}$ and $\mathcal{M_\mr{A}}$ should be given greater consideration since the fundamental physical processes at play are better reflected in such relative (dimensionless) quantities. The mass-to-flux ratio $\mu_{\mr{B}}$ is different in the three models because of the adjustments in $\rho_0$ and $B_0$, but the variations are not significant enough to affect the results. Tab.~\ref{tab:init} lists initial values associated with important quantities in each model.

It is to be noted that changing $\als$ without affecting any of the other dimensionless cloud parameters is not possible as many of them are mathematically and physically related. We choose to adjust $\als$ by changing the mean gas density, which also changes the mean thermal Jeans mass (and the number of Jeans masses) but keeps the other crucial parameters like $\mathcal{M}$ and $\mathcal{M_{\mr{A}}}$ fixed (see Tab.~\ref{tab:init}). Alternatively, one could adjust $\als$ by changing the velocity dispersion or the cloud size. However, changing the velocity dispersion would affect the Mach number, which would have distinct consequences on the cloud evolution (see \S\ref{sec:discussion}). Changing the cloud size would vary the number of Jeans masses too. Moreover, since $\als \propto L^{-2}$, the cloud size has to be increased to obtain lower virial parameters, which would make performing the corresponding simulations challenging due to the increase in the computational volume and therefore the computational expense.

\section{Results}
\label{sec:results}

\begin{table*}
\caption{Initial conditions and important results for different simulation models.}
\renewcommand{\arraystretch}{1.0}
\setlength{\tabcolsep}{3.0pt}
\label{tab:init}
\begin{tabular}{lrrrrrrrrrrrrrr}
\hline
\hline
 Model & $L\, [\mr{pc}]$ & $\rho_0\, [\mr{g\, cm^{-3}}]$ & $M_{\mr{cl}}\, [\mr{M_\odot}]$ & $M_{\mr{cl}}/M_{\mr{J}}$ & $\mathcal{M}$ & $B_0\, [\mr{\mu G}]$ & $\mathcal{M_{\mr{A}}}$ & $\mu_{\mr{B}}$ & $\beta$ & $N_{\mr{sims}}$ & $N_{\mr{tot}}$ & $\overline{M}_{\mr{median}}\,[\mathrm{M_\odot}]$ & $\overline{M}_{\mr{average}}\,[\mathrm{M_\odot}]$ & $\overline{M}_{50}\,[\mathrm{M_\odot}]$\\
 (1) & (2) & (3) & (4) & (5) & (6) & (7) & (8) & (9) & (10) & (11) & (12) & (13) & (14) & (15)\\
\hline
$\als = 0.0625$ & 2 & $5.25\times10^{-20}$ & 6200 & 2091.0 & 5 & 10 & 2.9 & 19.7 & 0.7 & 1 & 517 & $0.25\pm0.02$ & $0.50\pm0.05$ & $1.04\pm0.19$\\
$\als = 0.125$ & 2 & $2.62\times10^{-20}$ & 3100 & 737.1 & 5 & 28 & 2.9 & 13.7 & 0.7 & 2 & 520 & $0.27\pm0.05$ & $0.46\pm0.08$ & $0.93\pm0.15$\\
$\als = 0.5$ & 2 & $6.56\times10^{-21}$ & 775 & 92.4 & 5 & 20 & 2.9 & 6.9 & 0.7 & 10 & 449 & $0.46\pm0.06$ & $0.73\pm0.08$ & $1.38\pm0.16$\\
\hline
\hline
\end{tabular}
\\
\raggedright\textbf{Notes.} Column~1: MHD simulation models with different virial parameter $\als$. Column~2: size of the cloud/computational domain. Columns~3: initial mean gas density. Column~4: total mass of the cloud. Columns~5: number of Jeans masses in the cloud. Column~6: rms sonic Mach number. Column~7: initial magnetic field strength. Column~8: Alfv$\acute{\text{e}}$n Mach number. Column~9: mass-to-flux ratio. Column~10: plasma beta (thermal-to-magnetic pressure ratio). Column~11: number of simulations (random seeds for the turbulence driving) performed. Column~12: total number of sink particles formed. Column~13: median stellar mass. Column~14: mean stellar mass. Column~15: $M_{\mr{50}}$, which is the median mass in a cumulative mass function. Quantities in Columns~13-15 are averaged over the SFE range $1-5\%$ (represented by overbars).
\end{table*}

For $\als = 0.0625$, we carried out one simulation, which produced 517~sink particles (young stellar objects; YSOs). For $\als = 0.125$, we performed 2~simulations with different random seeds for the turbulence driving (Ornstein-Uhlenbeck process; see Sec.~\ref{sec:methods}, but otherwise identical cloud properties), which generated a total of 520~sink particles. In the case of $\als = 0.5$, we performed 10~simulations with different turbulence driving seeds, producing a total of 449~sink particles. In each of the models, the choice of the number of simulations performed is based on the aim to achieve a similar number of sink particles or similar IMF statistics among the three models. With the target of obtaining a statistically representative sample of the IMF, 10~simulations (turbulence seeds) were needed to produce a significant number of sinks or star particles ($>400$), while for $\als=0.125$ and $0.0625$, only 2 and 1~simulation(s), respectively, were required to produce a similar overall number of sink particles. For $\als = 0.125$ and $0.5$, different quantities studied here are averaged (or compiled together) over the set of simulations in each of the models (2 and 10~simulations, respectively), while for $\als = 0.0625$ it is from the single simulation performed. 

\begin{figure*}
    \centering
    \includegraphics[width=\textwidth]{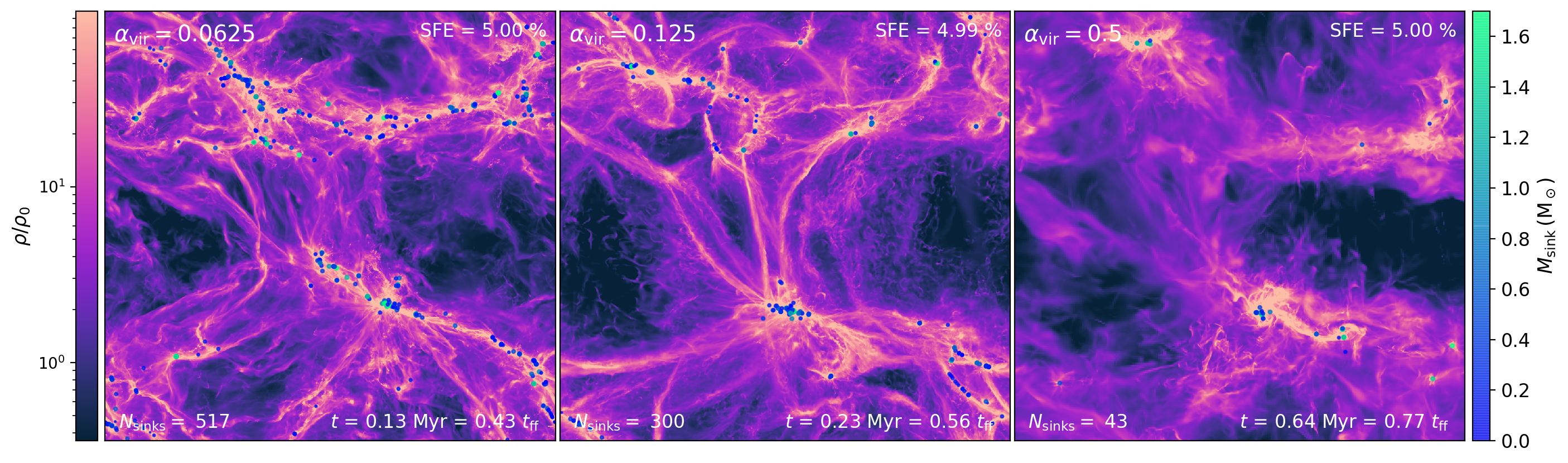}
    \caption{Mass-weighted gas density projection divided by the corresponding initial density for the models $\als = 0.0625$ (left panel), $\als = 0.125$ (middle panel), and $\als = 0.5$ (right panel) at star formation efficiency (SFE) of 5\%. The circular markers represent sink particle (star+disc system) positions and the colour bar on the right corresponds to the mass of the sink particles. See \S\ref{sec:app} for 3D visualisation of the respective simulations.} 
    \label{fig:densmap}
\end{figure*}

Fig.~\ref{fig:densmap} shows the mass-weighted gas density projection for the three simulation models at $\mathrm{SFE}=5\%$. For $\als = 0.5$ (right panel), the high-density structures are localised and there are mainly three star-forming regions, while in the lower virial parameter cases of $\als = 0.0625$ (left panel) and $\als = 0.125$ (middle panel), the high-density regions are spread across the simulation box with a large network of star-forming regions. Since gravity is more dominant in the lower virial parameter models, it is able to amplify the anisotropies produced by turbulence more strongly, creating additional sheet-like and filamentary structures \citep{vazquezsemadeni2007,2008MNRAS.389.1556B,2015MNRAS.450.4035F,2015ApJ...801...77M,vazquezsemadeni2019,2024arXiv240407301H}. Thus, a higher fraction of the cloud can form stars. 3D visualisations of the respective simulations are shown in \S\ref{sec:app}.

\begin{figure*}
    \centering
    \includegraphics[width=\textwidth]{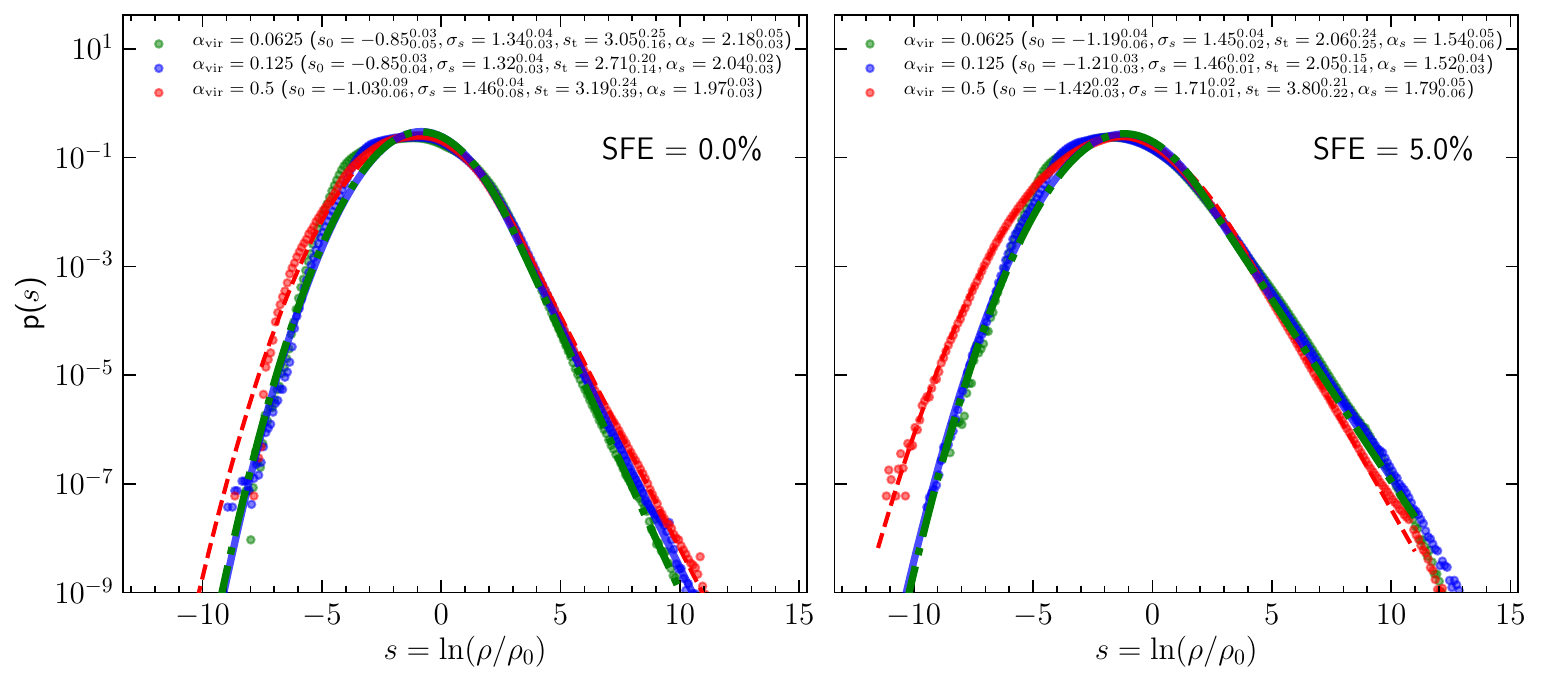}
    \caption{The volume-weighted gas density distribution for the three models in units of the corresponding initial density at the time the first sink particle forms (left panel) and at SFE = 5\% (right panel). A log-normal+power-law curve (see Eq.~\ref{Eq: denspdf_fit}) is fitted to the distributions, and the parameters of the fit, namely the peak and width of the log-normal part ($s_0$ and $\sigma_s$, respectively), the transition density between the log-normal and power-law parts ($s_\mr{t}$), and the slope of the power-law part ($\alpha_s$), are shown in the legends.} 
    \label{fig:denspdfs}
\end{figure*}

\subsection{Gas density probability distribution function}
Previous theoretical and numerical works find that the form of the density probability distribution function (PDF) of supersonic turbulence is approximately log-normal \citep{1994ApJ...423..681V,1997MNRAS.288..145P,2007ApJ...665..416K,2008ApJ...688L..79F,2013MNRAS.436.1245F,hopkins2013,2015MNRAS.448.3297F}. In the presence of gravity, the PDF develops a power-law tail in the high-density regime \citep{2000ApJ...535..869K,Kritsuk2011,Ballesteros-Paredes2011,collins2012,2013ApJ...763...51F,girichidis2014,schneider2015,Myers2015,2015ApJ...801...77M,mocz2017,padoan2017,Burkhart2018}. With time and increasing SFE, the slope of the power-law part (in logarithmic scale) becomes shallower \citep{2013ApJ...763...51F}, and the density at which the PDF transitions from log-normal to a power-law decreases \citep{Burkhart2018}. 

Fig.~\ref{fig:denspdfs} shows the PDFs of the volume-weighted logarithmic density contrast $s = \mr{ln}(\rho/\rho_0)$ in our simulations with $\alpha_{\mr{vir}} = 0.0625$ (green, dash-dotted), $0.125$ (blue, solid), and $0.5$ (red, dashed) at the moment the first sink particle forms (left panel) and at SFE = 5\% (right panel). The PDFs obtained from the simulations are plotted with circular markers and the curves denote the corresponding fits. Following the method in \citet{khullar2021}, we fit a log-normal+power-law function to the density PDFs, where the free parameters are the width of the log-normal part ($\sigma_s$) and the slope of the power-law part ($\alpha_s$). The form of the fit function is given by
 \begin{equation}
    \mr{p}(s) =
      \begin{cases}
        \frac{N_\mr{d}}{\sqrt{2\pi\sigma_s^2}}\,\exp\left(-\frac{(s - s_0)^2}{2 \sigma_s^2}\right)   & \text{for\, $s < s_\mr{t}$,}\\ 
        N_\mr{d}k_\mr{d}\, \exp\left(-s\alpha_s\right)  & \text{for\, $s \geq s_\mr{t}$.}\\
      \end{cases}
      \label{Eq: denspdf_fit}
\end{equation}
For defining the parameters like the peak of the log-normal part ($s_0$), transition density ($s_\mr{t}$), and the normalisation constants ($N_\mr{d}$ and $k_\mr{d}$), we use the corresponding expressions derived in \citet{khullar2021} based on constraints like continuity and differentiability at the transition density, mass conservation, and normalisation of the PDF to unity \citep[see \S2.2 in][for more details]{khullar2021}. The derived parameter values of the fitted function are shown in the legend of Fig.~\ref{fig:denspdfs}.  

\begin{figure}
    \centering
    \includegraphics[width=\columnwidth]{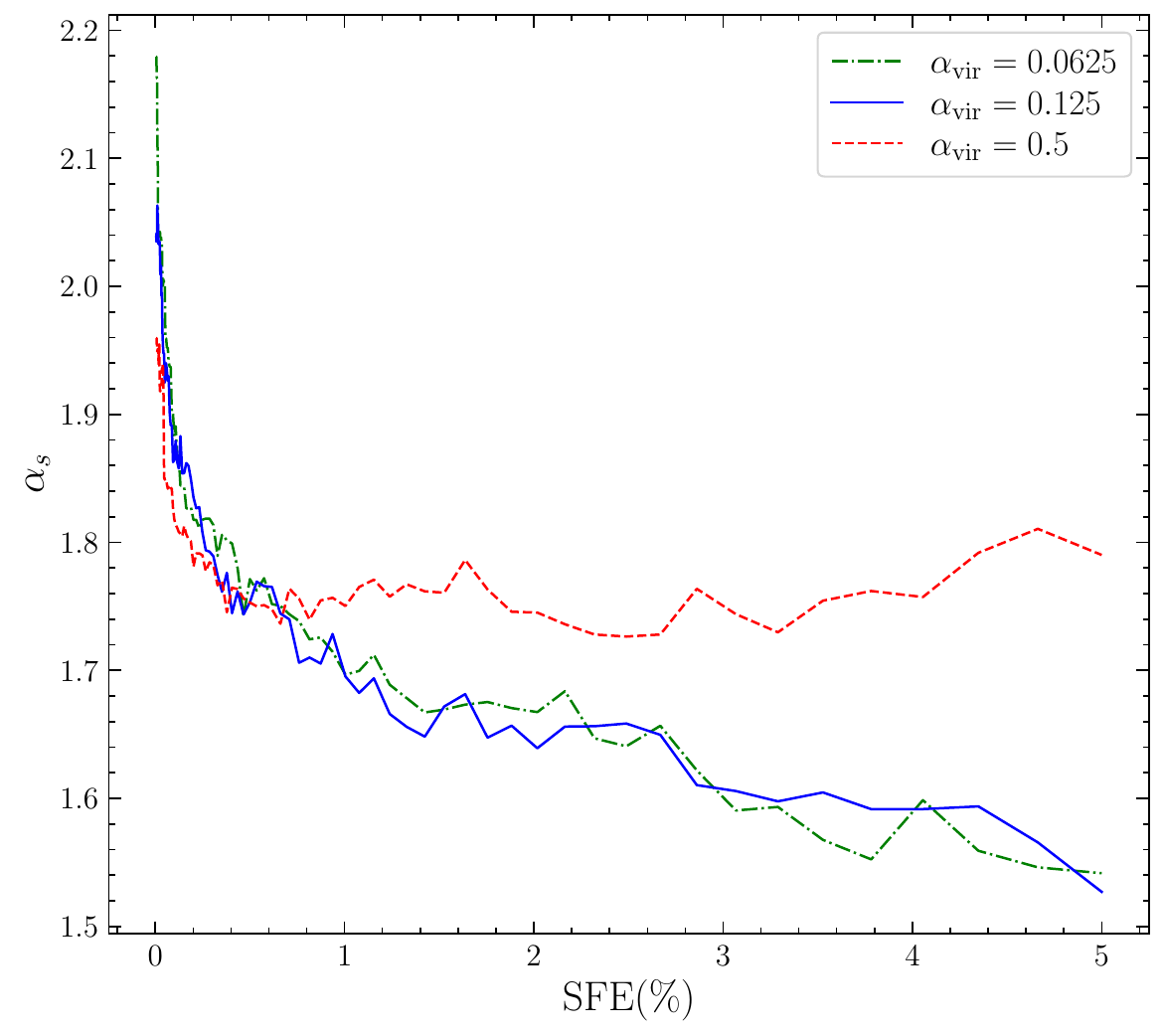}
    \caption{The slope of the power-law part of the density PDF as a function of SFE for $\als = 0.0625$ (green, dash-dotted), $0.125$ (blue, solid), and $0.5$ (red, dashed).}  
    \label{fig:alpha_fit}
\end{figure}

At the time the first star particle forms (SFE = 0\%; left panel), the density PDF for the $\alpha_{\mr{vir}} = 0.5$ model has a slightly higher fraction of dense gas (high $s$) than those with $\alpha_{\mr{vir}} = 0.0625$ and $\alpha_{\mr{vir}} = 0.125$. The reason for the relatively lower fraction in the simulations with $\alpha_{\mr{vir}} = 0.0625$ and $\alpha_{\mr{vir}} = 0.125$ is that stars start to form much earlier in these simulations (at $t\sim0.05-0.1\, \mr{Myr}$) compared to those with $\alpha_{\mr{vir}} = 0.5$, where the first stars form at $t\sim0.2\, \mr{Myr}$. Therefore, the clouds corresponding to $\alpha_{\mr{vir}} = 0.0625$ and $\alpha_{\mr{vir}} = 0.125$ are comparatively less dynamically evolved when compared at the moment the first star forms. Conversely, at SFE = 5\% (right panel), the simulations with $\alpha_{\mr{vir}} = 0.0625$ and $\alpha_{\mr{vir}} = 0.125$ have a higher fraction of high-$s$ regions than that for $\alpha_{\mr{vir}} = 0.5$. This is also apparent when comparing the slope of the power-law part $\alpha_s$ of the density PDF between the simulation models (compare legends in Fig.~\ref{fig:denspdfs}). In the left panel, $\alpha_s$ in the model with $\alpha_{\mr{vir}} = 0.5$ is shallower than that of the models with $\alpha_{\mr{vir}} = 0.0625$ and $\alpha_{\mr{vir}} = 0.125$, while in the right panel, it is steeper.

Fig.~\ref{fig:alpha_fit} shows the dependence of $\alpha_s$ on SFE. It is evident that $\alpha_s$ remains almost constant with time (SFE) in the case of $\als = 0.5$, which agrees with \citet{khullar2021} who find the same for $\als=0.5-2$ \citep[also, see][]{2022ApJ...927...75A,2023ApJ...954...93A}. This is because, in the case of high virial parameters, stars primarily form in the high-density regions initially seeded by turbulence, i.e., the mass reservoir is predetermined by the initial turbulence properties. On the other hand, $\alpha_s$ decreases (slopes become shallower) with SFE for $\als = 0.0625$ and $0.125$, which is because low virial parameters imply that gravity is more efficient in enhancing the dense structures in the cloud. Once gravity has had sufficient time to amplify the anisotropies and develop filamentary channels, simulations with lower virial parameters develop over-densities, and consequently form stars at a higher rate, which is discussed next.

\subsection{Star formation rate}
\label{sec:sfr_chap3}
The star formation rate is given by $\mr{SFR} = M_*/\tau_{\mr{c}}$, where $M_*$ is the cloud mass converted into stellar mass and $\tau_{\mr{c}}$ is the average time scale of the conversion process. SFR is generally referenced in its dimensionless form, i.e., in mass fraction per freefall time, given by \citep{sfr_km2005}
\begin{equation}
    \begin{split}
    \mr{SFR_{ff}} & = \frac{\mr{SFR}}{M_{\mr{cl}}}\times t_{\mr{ff,\, \rho_0}} \\
    &= \frac{M_*}{M_{\mr{cl}}}\times\frac{t_{\mr{ff,\, \rho_0}}}{\tau_{\mr{c}}},
    \end{split}
    \label{eq: sfr}
\end{equation}
where $M_{\mr{cl}}$ and $t_{\mr{ff,\, \rho_0}}$ are the total mass of the cloud and the freefall time at the mean density $\rho_0$, respectively.

\begin{figure}
    \centering
    \includegraphics[width=\columnwidth]{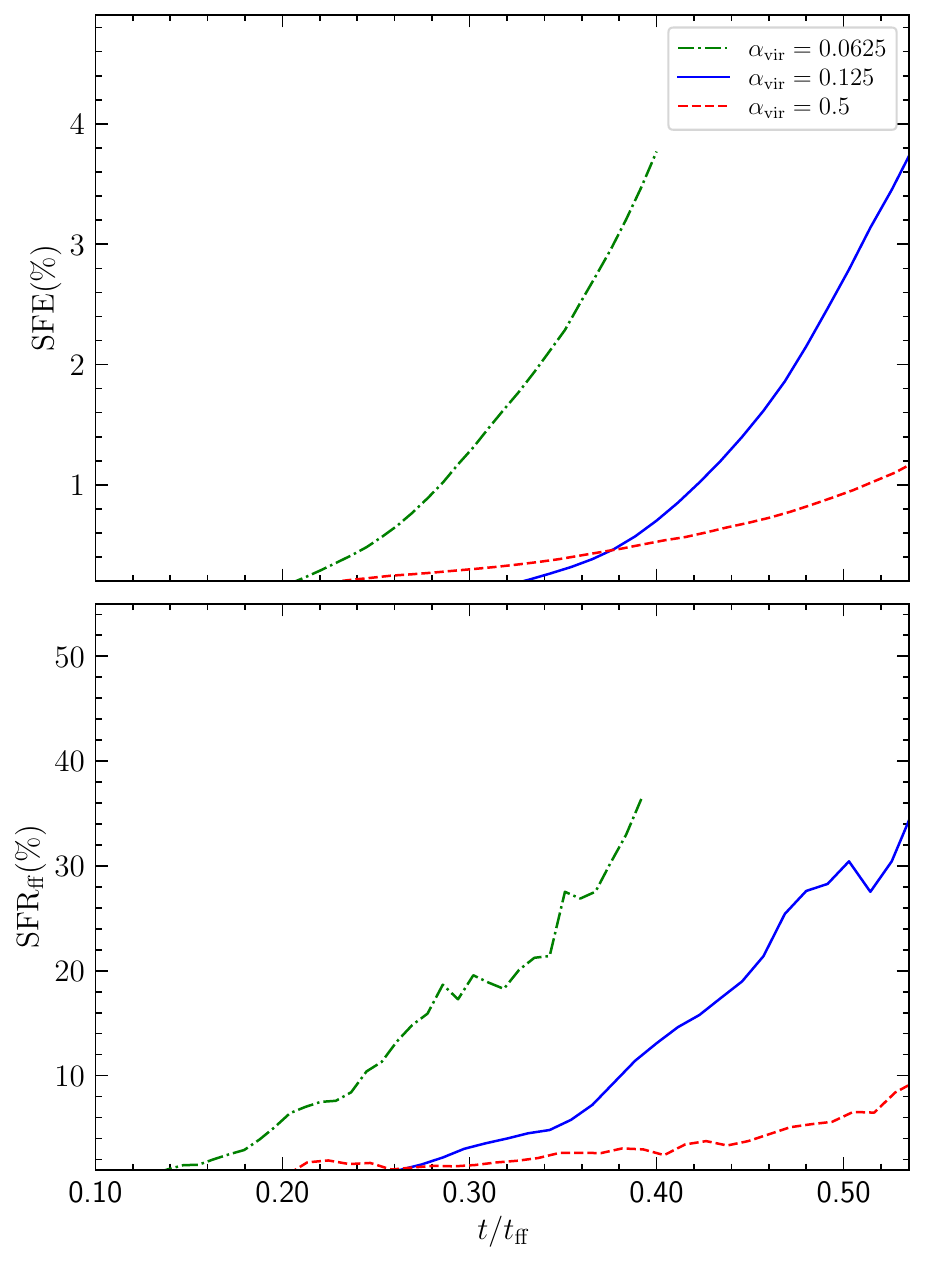}
    \caption{Top panel: The star formation efficiency $\mr{SFE}$ (\%), as a function of time (in units of the mean freefall time) for $\als = 0.0625$ (green, dash-dotted), $0.125$ (blue, solid), and $0.5$ (red, dashed). Note that the mean freefall time is different for each of the simulation models since their mean densities are different. Bottom panel: Same as the top panel, but for the star formation rate per freefall time, $\mr{SFR_{ff}}$.} 
    \label{fig:sfr}
\end{figure}

\begin{figure}
    \centering
    \includegraphics[width=\columnwidth]{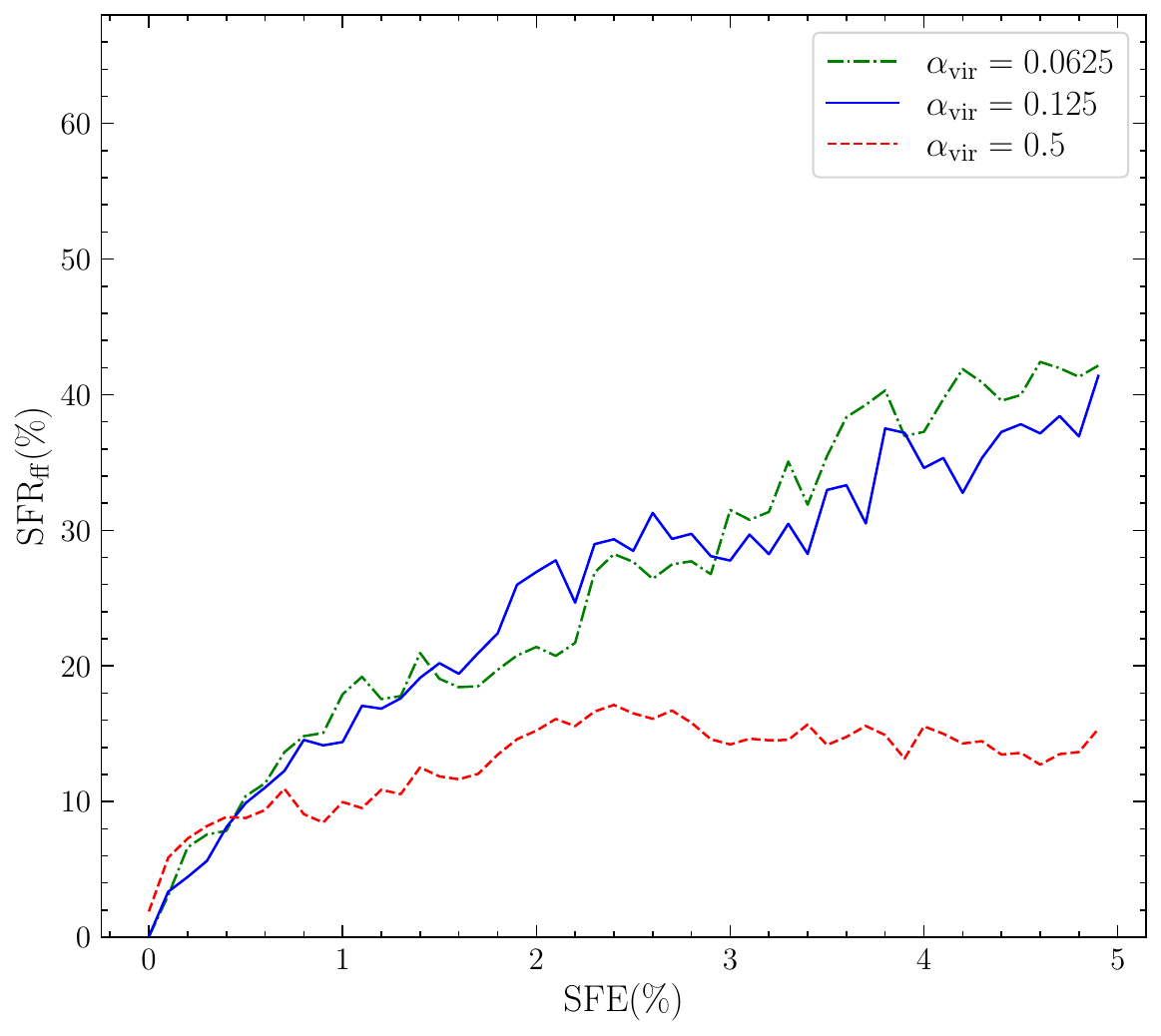}
    \caption{$\mr{SFR_{ff}}$ as a function of SFE (\%) in our simulations with different $\als$, showing that lowering $\als$ leads to an increase in $\mr{SFR_{ff}}$, but only mildly so when $\als\ll1$. This is due to a saturation of the effect of gravitational binding at very low $\als$.}
    \label{fig:sfr_bm}
\end{figure}

Fig.~\ref{fig:sfr} shows SFE (top panel) and $\mr{SFR_{ff}}$ (bottom panel) as a function of time in our simulations with $\als = 0.0625$ (green, dash-dotted), $0.125$ (blue, solid), and $0.5$ (red, dashed). It is clear that, when considered at the same time, $\mr{SFR_{ff}}$ is higher for lower virial parameters, with the highest $\mr{SFR_{ff}}$ in the simulation with $\als = 0.0625$. The offset between the $\mr{SFR_{ff}}$ curves for $\alpha_\mr{vir} = 0.0625$ and $\alpha_\mr{vir} = 0.125$ is simply because the onset of star formation is earlier in $\alpha_\mr{vir} = 0.0625$ than for $\alpha_\mr{vir} = 0.125$. The lower the value of $\als$, the faster gravity acts in rearranging gas to form stars. However, once the star formation gets underway, the rate of change of $\mr{SFR_{ff}}$ is similar for $\als = 0.0625$ and $0.125$, which becomes more evident when $\mr{SFR_{ff}}$ is compared at the same SFE. Fig.~\ref{fig:sfr_bm} shows the evolution of $\mr{SFR_{ff}}$ with SFE in our three simulation models. Here the $\mr{SFR_{ff}}$ values for $\als = 0.0625$ and $0.125$ are similar, while $\mr{SFR_{ff}}$ for $\als = 0.5$ is lower. When $\als$ is sufficiently low, the process is more strongly dictated by the gravitational interactions than other mechanisms and the SFR accelerates with time. In other words, when the virial parameter is sufficiently low, SFR is almost independent of the value of $\als$. Hence the rate of change of SFR is similar for $\alpha_\mr{vir} = 0.0625$ and $\alpha_\mr{vir} = 0.125$. However, in the case of $\alpha_\mr{vir} = 0.5$, the turbulent support still plays a substantial role throughout the process, and thus the acceleration in SFR is not as significant. The $\mr{SFR_{ff}}=0.30$, 0.29, and 0.14, averaged over the SFE range $1-5\%$, for $\als = 0.0625, 0.125$, and $0.5$, respectively.

\subsection{Initial mass function}

\label{sec:imf}
\begin{figure}
    \centering
    \includegraphics[width=\columnwidth]{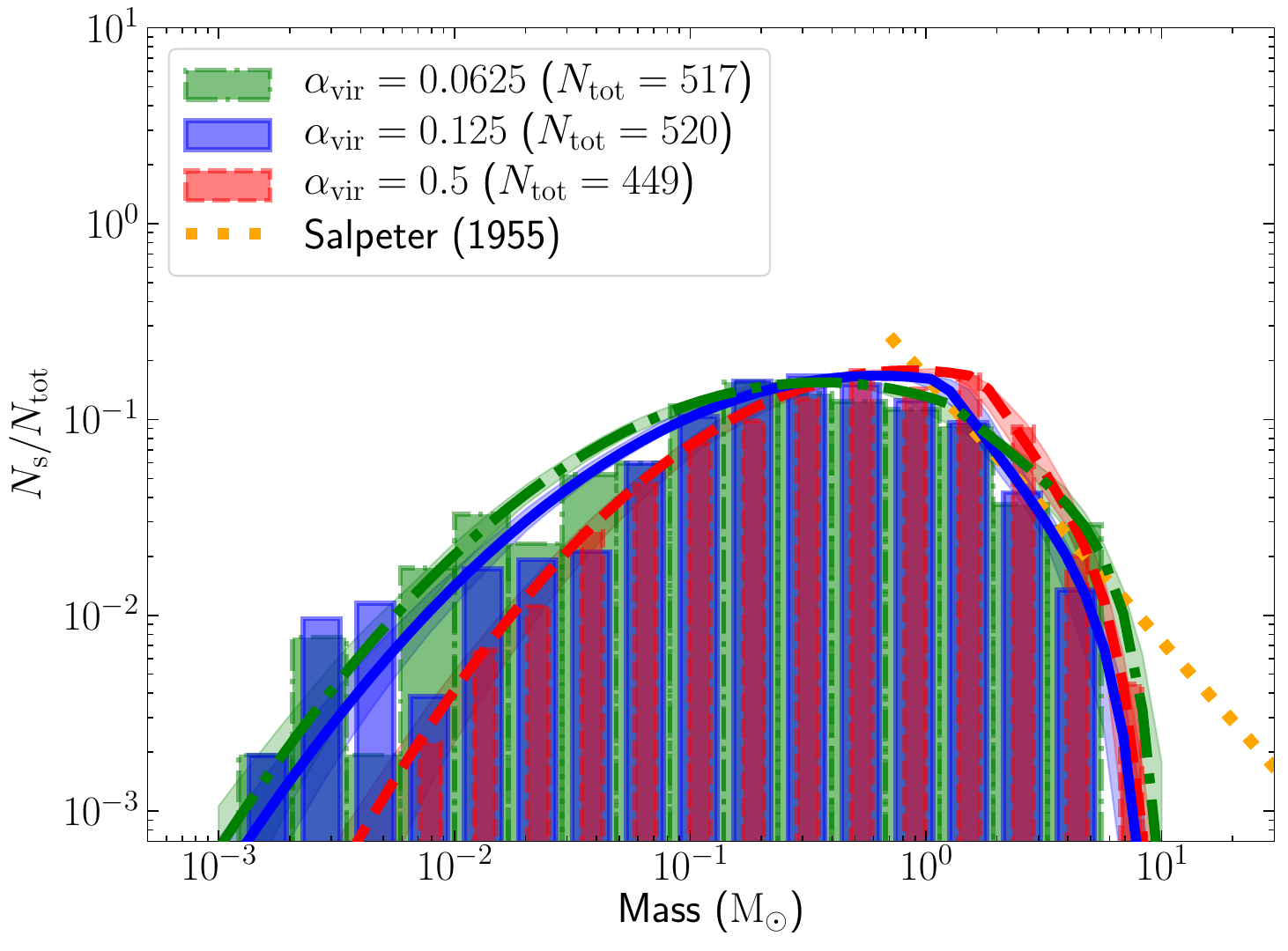}
    \caption{The IMF for simulations with $\als = 0.0625$ (green histogram with dash-dotted edges), $0.0125$ (blue histogram with solid edges), and $0.5$ (red histogram with dashed edges) at SFE = 5\%. Each bin represents the ratio of the number of star particles in the associated mass range ($N_{\mr{s}}$) to the total number of star particles ($N_{\mr{tot}}$). The dash-dotted, solid, and dashed curves represent the modified \citet{chabrier2005} IMF fits using Eq.~(\ref{Eq: Chab}) for the simulation IMFs corresponding to the models with $\als = 0.0625, 0.125$, and $0.5$, respectively (see \S\ref{sec:imf}). The parameters for the fit are derived using MCMC sampling and are listed in Tab.~\ref{tab:mcmc}. The dotted line shows the \citet{1955ApJ...121..161S} IMF.}  
    \label{fig:imf}
\end{figure}

Fig.~\ref{fig:imf} presents the mass distribution of sink particles (star+disc systems) obtained in the simulations with $\als = 0.0625$ (green histogram with dash-dotted edges), $0.0125$ (blue histogram with solid edges), and $0.5$ (red histogram with dashed edges). The curves represent a modified version of the \citet{chabrier2005} IMF fitted to the simulation data using the Markov Chain Monte-Carlo (MCMC) sampler emcee of \citet{2013PASP..125..306F} \citep[see also][]{2021MNRAS.503.1138N}. The input to the sampling algorithm is the list of sink particle masses rather than the binned data, and hence the method has the advantage that the fit is not affected by the binning choice. The modified version is similar to the standard \citet{chabrier2005} IMF form except that, to take into consideration the finite mass in our computational domain, we include an exponential term that serves as a smooth cutoff in the high-mass regime of the \citet{chabrier2005} IMF,
 \begin{equation}
    dN / d\mathrm{log} M =
      \begin{cases}
        k_1\exp\left(-\frac{1}{2}\left(\frac{\log M-\log M_0}{\sigma}\right)^2\right)   & \text{for\, $M < M_{\mathrm{T}}$,}\\ 
        k_2\,M^{-\Gamma}\exp\left(-(M/M_{\mathrm{cut}})^{p}\right)  & \text{for\, $M \geq M_{\mathrm{T}}$,}\\
      \end{cases}
      \label{Eq: Chab}
\end{equation}
where all masses are in units of $\mathrm{M_\odot}$. Here the appropriate values for the five free parameters $\theta_{\mr{fit}} = (\sigma, \mathrm{log}\, M_0, \mathrm{log}\, M_{\mathrm{T}}, \Gamma, \mathrm{log}\, M_{\mathrm{cut}})$ are derived using MCMC sampling based on the simulated data. $\sigma, M_0, M_{\mathrm{T}}$ and $\Gamma$ are the standard deviation of the log-normal part, the peak mass, transition mass between the log-normal and power-law forms, and the slope of the power-law part, respectively. $k_1$ and $k_2$ are normalisation factors, defined to ensure continuity at $M_{\mathrm{T}}$. As a consequence of the exponential term in the power-law part, the IMF will be truncated at high masses. $M_{\mathrm{cut}}$ characterises the mass at which the exponential term begins to dominate and $p$ describes how sharply the cut-off occurs around $M_{\mathrm{cut}}$. We use $p = 4$ to achieve a sufficiently sharp drop around $M_{\mathrm{cut}}$. We refer the reader to \S3.2 in \citet{mathew2023} for a more detailed description of the MCMC fitting technique used here.

\begin{table*}
\caption{Characteristics of the sink particle mass distribution and parameter values from the MCMC fit to the distribution.}
\renewcommand{\arraystretch}{1.5}
\label{tab:mcmc}
\begin{tabular}{ccccccccccc}
\hline
\hline
 \multirow{2}{*}{Model} & & \multicolumn{2}{c}{Simulation} & & \multicolumn{6}{c}{MCMC fit} \\
 \cline{3-4} \cline{6-11}
 & & Peak $[\mathrm{M_\odot}]$ & Median [$\mathrm{M_\odot}$] & & $M_0\, [\mathrm{M_\odot}]$ & $\sigma$ & $M_{\mathrm{T}}\, [\mathrm{M_\odot}]$ & $\Gamma$ & $M_{\mathrm{cut}}\, [\mathrm{M_\odot}]$ & $p$ \\
 (1) & & (2) & (3) & & (4) & (5) & (6) & (7) & (8) & (9) \\
\hline
$\als = 0.0625$ & & 0.14-0.24 & $0.27\pm0.02$ & &  $0.36_{-0.07}^{+0.13}$ & $0.77_{-0.06}^{+0.08}$ & $1.25_{-0.53}^{+2.35}$ & $0.90_{-0.30}^{+0.60}$ & $7.10_{-1.20}^{+1.30}$ & 4 \\
$\als = 0.125$ & & 0.24-0.40 & $0.33\pm0.02$ & & $0.63_{-0.15}^{+0.19}$ & $0.81_{-0.06}^{+0.06}$ & $1.17_{-0.28}^{+0.44}$ & $1.60_{-0.40}^{+0.70}$ & $6.30_{-1.10}^{+1.30}$ & 4 \\
$\als = 0.5$ & & 0.40-0.67 & $0.54\pm0.04$ & & $0.86_{-0.20}^{+0.29}$ & $0.71_{-0.06}^{+0.07}$ & $1.70_{-0.33}^{+0.84}$ & $1.70_{-0.40}^{+0.90}$ & $6.50_{-0.90}^{+1.20}$ & 4 \\
\hline
\hline
\end{tabular}
\\
\raggedright\textbf{Notes.} Column~1: MHD simulation models with different virial parameter $\als$. Column~2: peak in the sink mass distribution (see Fig.~\ref{fig:imf}). Columns~3: median of the mass distribution. Columns~4-9: $50^{\mathrm{th}}$ percentile of the IMF parameters derived from the MCMC fit to the distribution, with the uncertainty bracketed by the $16^{\mathrm{th}}$ and $84^{\mathrm{th}}$ percentiles. All quantities presented here represent the measurement at SFE = 5\%.
\end{table*}

The three stellar mass distributions are not substantially different, but they exhibit some distinctions. The derived values for the fit parameters corresponding to the simulated IMFs are listed in Tab.~\ref{tab:mcmc}. The distributions for the lower virial parameters have a higher fraction of low-mass stars, which is expected since the large-scale turbulent support is weaker (or equivalently, gravity is comparatively more efficient), and therefore there is more fragmentation. The peak of the distributions for $\als = 0.0625$ and $\als = 0.125$ are comparable, however, the peak for $\als = 0.5$ is at a relatively higher mass (see Tab.~\ref{tab:mcmc}). Changing the virial parameter in the simulation from $\als = 0.5$ to $\als = 0.0625$, i.e., by a factor of 8, reduces the peak mass by a factor of $\sim2.4$. This shows that the IMF has a relatively weak, but systematic dependence on the cloud virial parameter. Lower $\als$ suggests that the overdensities are more bound and stable against disruption by shocks \citep{sfr_fk2012,2016MNRAS.462.4171B}, and hence more low-mass overdensities can collapse to form stars. 

The slopes $\Gamma$ of the fit at the high-mass end for $\als = 0.0625, 0.125,$ and $0.5$ are $0.90_{-0.30}^{+0.60}, 1.60_{-0.40}^{+0.70},$ and $1.70_{-0.40}^{+0.90}$, respectively (note that the Salpeter slope would be $\Gamma=1.35$ in this definition of $\Gamma$). The slope for $\als = 0.0625$ is shallower than that of the other $\als$ values. However, since the high-mass range is narrow in our simulations, the error bars on the $\Gamma$ estimates are large and therefore it is difficult to make conclusive remarks on the dependence of the high-mass regime of the IMF on the cloud virial parameter just based on $\Gamma$.     

\begin{figure}
    \centering
    \includegraphics[width=\columnwidth]{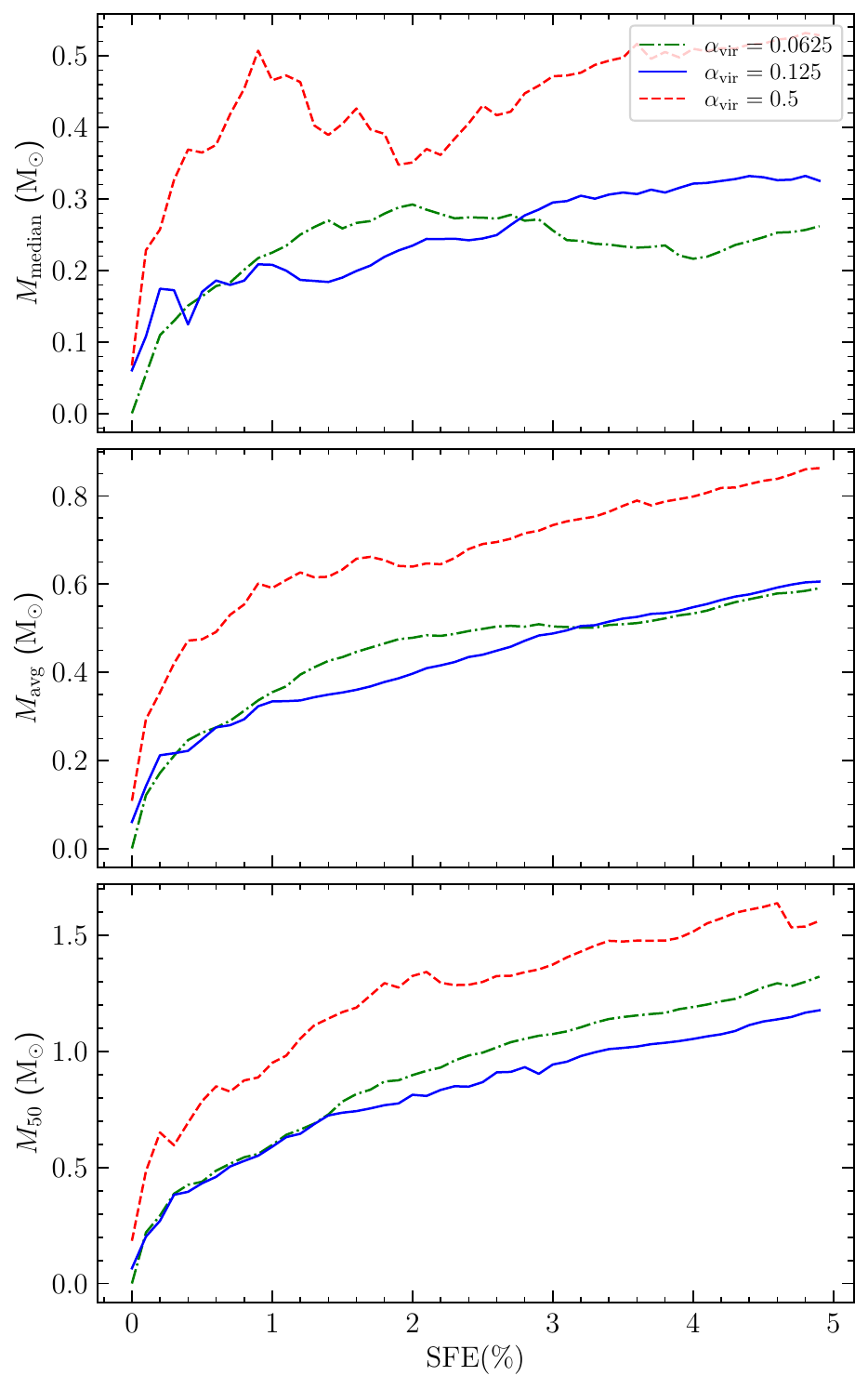}
    \caption{Top panel: The median stellar mass as a function of SFE (\%) for $\als = 0.0625$ (green, dash-dotted), $0.125$ (blue, solid), and $0.5$ (red, dashed). Middle panel: same as top panel, but for the average stellar mass. Bottom panel: same as top panel, but for $M_{50}$, which is the median mass in a cumulative mass function.}  
    \label{fig:avg_mass}
\end{figure}

The median masses of the IMFs at SFE = 5\% for $\als = 0.0625, 0.125,$ and $0.5$ are $\sim 0.26\,\mathrm{M_\odot}, 0.32\,\mathrm{M_\odot}$, and $0.53\, \mathrm{M_\odot}$, respectively. The average masses at SFE = 5\% are $\sim 0.59\,\mathrm{M_\odot}, 0.60\,\mathrm{M_\odot}$, and $0.86\, \mathrm{M_\odot}$, respectively. We also evaluate $M_{\mathrm{50}}$, which is, as outlined in \citet{2012ApJ...754...71K}, the median mass in a cumulative mass distribution, i.e., $M_{\mathrm{50}}$ is the value below which 50\% of the total stellar mass is found. We find $M_{\mathrm{50}} \sim 1.32\,\mathrm{M_\odot}, 1.17\,\mathrm{M_\odot}$, and $1.56 \, \mathrm{M_\odot}$ at SFE = 5\% for $\als = 0.0625, 0.125,$ and $0.5$, respectively. Fig.~\ref{fig:avg_mass} shows the evolution of the median mass (top panel), the average mass (middle panel), and $M_{50}$ (bottom panel) with SFE in the three $\als$ models. The median mass remains relatively constant from $\mr{SFE\sim1}$ for the three models, while the average mass increases moderately with SFE. The increase in the average mass reflects the development of the high-mass end of the IMF with time. The median, mean and $M_{50}$ masses averaged over the SFE range $1-5\%$ for $\als = 0.0625, 0.125,$ and $0.5$ are listed in Tab.~\ref{tab:init}.

\subsection{Comparison with observational IMF}
\begin{figure*}
    \centering
    \includegraphics[width=2\columnwidth]{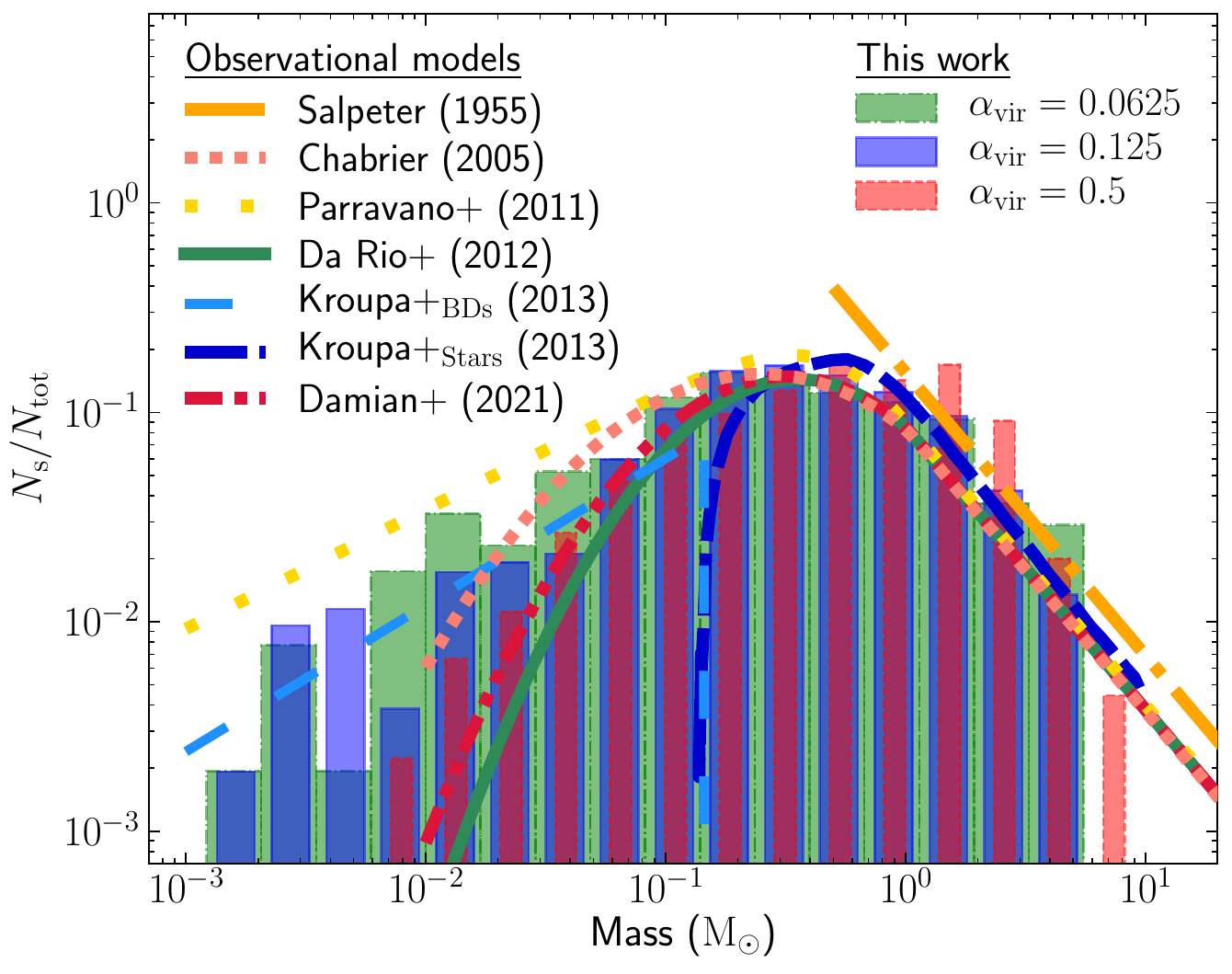}
    \caption{Comparison of our simulation IMFs at SFE = 5\% with the different IMF models based on observational surveys. The plotted curves are the system IMF models by \citet{1955ApJ...121..161S} (dash-dotted), \citet{chabrier2005} (short-dotted), \citet{2011ApJ...726...27P} (long-dotted), \citet{2012ApJ...748...14D} (solid), \citet{2013pss5.book..115K} for brown dwarfs (long-dashed) and stars (short-dashed), and \citet{2021MNRAS.504.2557D} (dash-dot-dotted).}  
    \label{fig:imf_obs}
\end{figure*}

Figure~\ref{fig:imf_obs} depicts the comparison between our simulation IMFs and the observational IMF models. We compare our data with the system IMFs (unresolved close binaries) rather than the individual-star IMF since the accretion discs in our simulations are not fully resolved. All the observational models broadly agree with each other, though there are slight variations in the low-mass regime. While \citet{2011ApJ...726...27P} propose a higher fraction of very low-mass stars or brown dwarfs than the \citet{chabrier2005} IMF, \citet{2012ApJ...748...14D} and \citet{2021MNRAS.504.2557D} suggest a lower fraction than \citet{chabrier2005}. Such a difference also exists between the IMFs for our three simulation models, where the simulation corresponding to $\als = 0.0625$ has the highest fraction of very low-mass objects and the simulations with $\als = 0.5$ have the lowest fraction. 

For comparisons between observations and simulations, it is also beneficial to measure the ratio between the number of stars in different mass ranges, which is more robust than deriving quantities such as the peak mass or the power-law slope. Observations find that the ratio of the number of sub-stellar objects ($M \leq 0.08\, \mathrm{M_\odot}$) to that with stellar masses ($0.15 < M \le 1.0\, \mathrm{M_\odot}$) is $\sim 0.2$, i.e., one brown dwarf (sub-stellar) is formed per every five late-type (sub-solar) stars on average \citep{2006AJ....132.2296A,2008ApJ...683L.183A,2007ApJ...671..767T,2011ApJ...726...27P,2013pss5.book..115K}. The corresponding ratios for the simulation models with $\als = 0.0625, 0.125,$ and $0.5$ are $0.41, 0.25$, and $0.19$, respectively. The ratio for $\als = 0.5$ agrees with the typical value from observations, while that for $\als = 0.0625$ is higher by a factor of $\sim2$. 

\subsection{Multiplicity}
\label{sec: multiplicity}
The multiplicity statistics is closely linked to the IMF and is a critical probe to test star formation theories. We identify multiple systems in our simulations based on the procedure employed in \citet{2009MNRAS.392..590B}, which we briefly summarise. In the list of stars (single objects) obtained from the simulations, we search for the closest gravitationally bound pair. The corresponding two stars are removed from the list and replaced by a binary object with the mass, centre-of-mass position and centre-of-mass velocity of the removed bound pair. In the new list, we again find the closest bound pair. In the scenario where the pair consists of a single and a binary object, the pair is substituted by a triple object. The above algorithm is repeated until no new bound pairs can be found in the list or the only possible option is a quintuple. We rule out pairings that result in a quintuple or systems of higher order since such high-order multiple systems tend to be dynamically unstable and are likely to decay to lower-order systems as the cloud evolves. This procedure is the same as the one used in \citet{mathew2021} and \citet{mathew2023}.

With the list of singles, binaries, triples and quadruples derived from the above technique, we calculate the multiplicity fraction in different mass ranges. The multiplicity fraction ($mf$) in a mass range is the ratio of the number of multiple systems to the total number of systems whose primary star falls within the specified mass range, i.e.,
\begin{equation}
    \label{eq:mf}
    mf =  \frac{B + T + Q}{S + B + T + Q},
\end{equation}
where $S, B, T$, and $Q$ represent the number of singles, binaries, triples, and quadruples, respectively.

\begin{figure}
    \centering
    \includegraphics[width=\columnwidth]{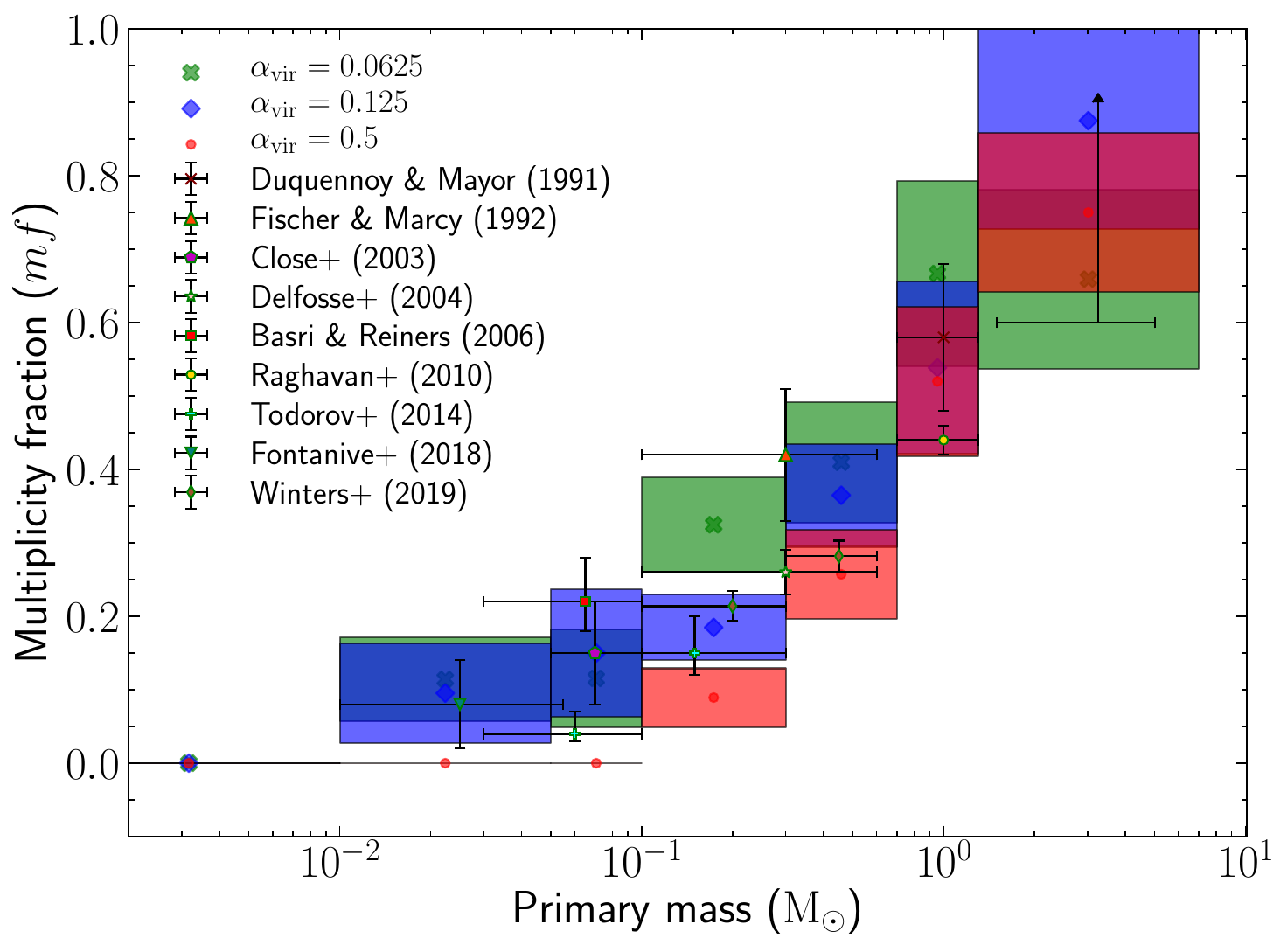}
    \caption{Multiplicity fraction ($mf$) computed via Eq.~(\ref{eq:mf}) in different primary mass intervals for the simulation models with $\als = 0.0625$ (green crossed markers and boxes), $\als = 0.125$ (blue diamond markers and boxes) and $\als = 0.5$ (red circular markers and boxes). The markers represent the $mf$ in the mass range denoted by the width of the box enclosing the marker. The height of the box represents the error margin in the $mf$ obtained. The multiplicity fractions measured in different observations are depicted by the centre of the crosses, with the horizontal and vertical components representing the mass interval considered in the survey and the uncertainties, respectively. The observational data are (from low to high primary mass), from \citet{2018MNRAS.479.2702F}, \citet{2014ApJ...788...40T}, \citet{2006AJ....132..663B}, \citet{2003ApJ...587..407C}, \citet{2014ApJ...788...40T}, \citet{2019AJ....157..216W} (not corrected for undetected companions), \citet{2004ASPC..318..166D}, \citet{1992ApJ...396..178F}, \citet{2010ApJS..190....1R} and \citet{1991A&A...248..485D}. The $mf$ for high-mass stars is not well understood. The lower limit of $mf$ in the mass range of $1.5$--$5\,\mathrm{M_\odot}$ is $\sim$ 0.5--0.6 \citep{2012MNRAS.424.1925C,2013ARA&A..51..269D}. Massive stars are considered to have $mf\sim1$ \citep{2009AJ....137.3358M,2011IAUS..272..474S,2017A&A...599L...9S,2020SSRv..216...70L}.}
    \label{fig:Multiplicityfraction}
\end{figure}

We show $mf$ as a function of the primary mass for the three simulation models in Fig.~\ref{fig:Multiplicityfraction}. We see that $mf$ increases with primary mass, which is consistent with the established understanding \citep{2012MNRAS.419.3115B,2012ApJ...754...71K,2013ARA&A..51..269D,2018MNRAS.476..771C,2020MNRAS.497..336S,offner2023}. It is evident that, at sub-solar masses, $mf$ for $\als = 0.0625$ and $\als = 0.125$ in each of the mass intervals are higher than that for $\als = 0.5$. The multiplicity fraction over the sub-solar range (primary mass $< 1\, \mr{M_\odot}$) for $\als = 0.0625, 0.125$, and $0.5$ are $\sim 0.27, 0.22,$ and $0.14$, respectively. The higher $mf$ in lower $\als$ cases might be because the local region in which the binaries form will be more bound and gas-rich in the case of low virial parameters. Therefore, the binaries are better shielded from dynamical interactions that result in binary decay \citep{2024ApJ...968...80R}. At super-solar masses, the derived $mf$ values are similar and $\gtrsim 0.6$ for the three cases. High-mass binaries generally form in the most gas-rich regions within the cloud, which is why they can accrete gas rapidly. Hence, the high-mass binaries are relatively shielded from interactions even in the highest virial parameter case of $\als = 0.5$. We note that, to confirm our interpretation of the higher $mf$ in lower $\als$ cases, a comprehensive quantitative study is required, measuring the local virial parameter in cluster regions and analysing the evolution of binary orbits, which is beyond the scope of this work. Finally, we find that the fraction of singles across the full mass range is $\sim 0.65$, irrespective of $\als$. Previous numerical and observational works also yield similar values for the fraction of single stars \citep{2006ApJ...640L..63L,2021MNRAS.500.3594R,mathew2021}.

\section{Discussion}
\label{sec:discussion}

\subsection{Comparison with previous numerical works}
\label{sec:dis_imf}
In our study, the mean thermal Jeans mass $M_{\mr{J}}$ and the number of Jeans masses for the three simulations models are different since the initial mean density varies between them. \citet{bateandbonnell2005} find that the IMF is linearly dependent on the mean thermal Jeans mass in their simulations. However, \citet{bate2009} establish that the inclusion of radiative feedback removes the dependence on the thermal Jeans mass, finding that the IMFs are indistinguishable for simulations with mean Jeans masses of $1\, \mr{M_\odot}$ and $1/3\, \mr{M_\odot}$ \citep[see also][]{2011ApJ...740...74K}. In our study, the mean Jeans mass in the simulations with $\als = 0.0625$ and $\als = 0.125$ also varies by a factor of $\sim 3$. However, contrary to our case, $\als$ is similar for the two simulation sets with different $M_{\mr{J}}$ in \citet{bate2009} as a result of using a different combination of cloud size and $\mathcal{M}$ in each of their two sets. Therefore, a direct comparison between the two studies cannot be made. 

\citet{2018A&A...611A..88L} performed a similar numerical experiment to \citet{bate2009} by changing the gas density (and hence the thermal Jeans mass) while keeping $\als$ fixed, by adjusting the cloud size and $\mathcal{M}$. They find that the IMF is relatively flat in the $0.1-1\, \mr{M_\odot}$ range at low densities, while at higher densities, the distribution starts to become more peaked, producing a power-law distribution with $\Gamma \sim 3/4$ at masses greater than a few $0.1\, \mr{M_\odot}$. However, stellar radiative heating was absent in their simulations, and therefore the influence of varying densities or thermal Jeans masses might be overestimated \citep{bate2009,2011ApJ...740...74K,2016MNRAS.458..673G,2017JPhCS.837a2007F,mathew2020,hennebelle2020}.

\citet{2015MNRAS.452..566B} and \citet{2016MNRAS.462.4171B} investigated the effect of the turbulent Mach number and found that the form of the IMF remains the same irrespective of $\mathcal{M}$ in their simulations. This contradicts the analytical theories of the CMF or the IMF, which predict that the characteristic mass decreases with an increase in the Mach number \citep{2002ApJ...576..870P,2008ApJ...684..395H,2012MNRAS.423.2037H}. According to these theories, a higher Mach number results in stronger shocks, leading to greater density contrasts and higher fragmentation. The observed trends in \citet{2015MNRAS.452..566B} and \citet{2016MNRAS.462.4171B} arise because raising the Mach number through the increase of the velocity dispersion also increases the cloud virial parameter, as $\als \propto \sigma_{\mr{v}}^2$. Higher virial parameters imply that over-densities are less susceptible to collapse, with only the sufficiently massive ones eventually collapsing \citep{2016MNRAS.462.4171B}. On the other hand, \citet{2018A&A...611A..88L} observe slight variations in the IMF on changing the Mach number in their simulations by a factor of $5$. Here again, the Mach number is varied by adjusting the velocity dispersion, which alters the virial parameter. 

\citet{guszejnov2022} find that a higher virial parameter results in a lower star formation rate, which agrees with our results. They find that variations in the virial parameter of the cloud do not significantly affect the form of the IMF, while in our study we see a weak dependence of the IMF on $\als$. However, \citet{guszejnov2022} also vary the virial parameter by adjusting the velocity dispersion, which in turn affects $\mathcal{M}$. Since a higher Mach number is expected to enhance the formation of low-mass over-densities and a higher virial parameter is expected to suppress them, the net impact on the IMF is expected to be minimal as seen in the simulations of \citet{2015MNRAS.452..566B}, \citet{2016MNRAS.462.4171B}, and \citet{2018A&A...611A..88L}. At very high Mach number (or very high $\sigma_\mr{v})$, the effect of the virial parameter (increased kinetic support) can dominate over the effect of the Mach number (density-enhancement via stronger shocks) and quench star formation altogether since $\als \propto \sigma_{\mr{v}}^2$, while $\mathcal{M} \propto \sigma_{\mr{v}}$. However, the details of the dependence are non-linear and depend on the region of parameter space considered \citep{sfr_fk2012}.

The numerical experiments carried out here are distinct from the numerical approaches discussed above. While the above works concentrate on the influence of the mean thermal Jeans mass (keeping the virial parameter constant) or the Mach number (keeping the mean thermal Jeans mass constant), i.e., in a sense the impact of the mean density or the velocity dispersion, we focus on the effect of the virial parameter, but keep the Mach number fixed. The mean density and hence the mean thermal Jeans mass varies between our simulation models, which could have some effect on the initial fragmentation in the cloud. However, since the turbulence in our simulations is supersonic ($\mathcal{M} = 5$), the kinetic energy dominates over the thermal energy and hence will govern most of the fragmentation in the early stages. Further, once the stars start to form, the protostellar heating increases the local temperature and Jeans mass in the star-forming cores on sub-parsec scales \citep{bate2009,2011ApJ...740...74K,2016MNRAS.458..673G,2017JPhCS.837a2007F,mathew2020,hennebelle2020}. Thus, the IMF variations that we observe between the three simulation models may be determined more by the variation in $\als$ than by variations in the mean thermal Jeans mass.

\citet{Haugb_lle_2018} carried out a numerical experiment similar to our analysis, where they changed the virial parameter by varying the mean density, but keeping $\mathcal{M}$ fixed. They find that lowering $\als$ shifts the peak of the IMF to lower masses, which aligns with our findings. However, \citet{Haugb_lle_2018} focus on a regime with $\als \sim 0.2-2$, while we concentrate on a lower $\als$ regime ($0.06-0.5$). \citet{Haugb_lle_2018} study the $\als$ range where either the gravitational and turbulent energies are comparable ($\als \sim 0.5$) or turbulence dominates ($\als\gtrsim1$), while we study the $\als$ range where either the gravitational and turbulent energies are comparable ($\als \sim 0.5$) or gravity dominates ($\als\lesssim0.1$). Thus, the present work is complementary to the works of \citet{bateandbonnell2005}, \citet{bate2009},  \citet{2015MNRAS.452..566B}, \citet{2016MNRAS.462.4171B}, \citet{2018A&A...611A..88L}, \citet{Haugb_lle_2018}, and \citet{guszejnov2022}. Ultimately, star formation may depend on a combination of all of the dimensionless parameters (e.g., $\mathcal{M}$, $\als$, $b$, $\mu_{\mr{B}}$, and $\mathcal{M_{\mr{A}}}$), and future work is needed to quantify the effects of each of these parameters on the SFR and IMF.

\subsection{Outlier IMFs}
The high gas density and large velocity dispersion measured in massive, elliptical early-type galaxies (ETGs) have been suggested as candidates for explaining their bottom-heavy nature \citep{2014ApJ...796...75C,2022A&A...666A.113D}. However, evaluation of the virial parameter in these regions is crucial to obtain conclusive results since understanding the respective $\als$ regime is important to comprehend the net effect of turbulence. Large velocity dispersion or very high $\als$ can quench star formation because the large-scale support by turbulence dominates over its ability to enhance fragmentation on small scales \citep{sfr_fk2012,2014ApJ...796...75C,brucy2024}. Another possible candidate that can partly contribute to the bottom-heavy nature is the mode of turbulence driving in such regions. Since ETGs are considered to be formed in a starburst event associated with galaxy mergers, the turbulence driving is likely dominated by compressive modes \citep{2008ApJ...688L..79F,2009ApJ...706...67R}, which can enhance the formation of low-mass stars \citep{2014ApJ...796...75C,mathew2023}. Further, our results suggest that the high virial parameters ($\als > 1$) measured in the Central Molecular Zone (CMZ) \citep{2022ApJ...929...34M} contribute in part to the low SFR and top-heavy IMFs \citep{1999ApJ...525..750F,2006ApJ...653L.113K,2013ApJ...764..155L,2019ApJ...870...44H} observed in the embedded clouds, with the prevalence of solenoidal driving \citep{2016ApJ...832..143F,2022MNRAS.515..271R} being another factor \citep[see also][]{2007MNRAS.374L..29K,Haugb_lle_2018,mathew2023}. We note that, given the non-linear IMF dependence found in our study, to understand the extent of the contribution of the virial parameter to the top-heavy nature of the IMF observed in the CMZ, regimes where $\als>1$ have to be considered, which are not directly explored in this study.

\subsection{Origin of the IMF peak}
The weak dependence of the IMF on the cloud properties like the virial parameter and mode of turbulence driving \citep{2002ApJ...576..870P,2008ApJ...684..395H,2010A&A...516A..25S,2012MNRAS.423.2037H,mathew2023} may account for why the region around the IMF peak is broad or resembles a plateau \citep[see also][]{2024arXiv241007311K}. When the net effect of the combination of cloud properties favours extensive fragmentation, the peak shifts to lower masses, and when the combination tends to suppress fragmentation, the peak shifts to higher masses. Nonetheless, it still needs to be addressed why the broad peak or plateau exists in the sub-solar or more specifically in the M-dwarf range. Some preferred candidates are the feedback mechanisms that emerge from the stars themselves, such as radiative heating and protostellar outflows, which have been found to be effective in self-regulation \citep{2006ApJ...640L.187L,bate2009,2010ApJ...709...27W,2011ApJ...740...74K,2014ApJ...790..128F,2014prpl.conf..451F,2016MNRAS.458..673G,2017ApJ...847..104O,2017JPhCS.837a2007F,mathew2020,hennebelle2020,mathew2021,2021MNRAS.502.3646G,2024A&A...683A..13L}, or generally, the thermodynamics in the immediate vicinity of stars \citep{2003ApJ...592..975L,2005MNRAS.359..211L,2005A&A...435..611J,2008ApJ...681..365E,2019ApJ...883..140H,2020MNRAS.492.4727C,2023arXiv230816268G}. 

\section{Conclusions}
\label{sec:conclusions}
We conduct a series of star cluster formation simulations to investigate the impact of the cloud virial parameter $\als$ on the formation process. The simulations incorporate gravity, magnetic fields, turbulence, protostellar radiative heating and mechanical feedback in the form of jets/outflows. We have three sets of simulations corresponding to $\als = 0.0625, 0.125$ and $\als = 0.5$. We observe that the number of stars formed increases with decreasing $\als$. In the case of high virial parameters, the formation of large-scale structures that host star clusters is primarily a consequence of turbulent shocks. By lowering the virial parameter, the relative influence of gravity in the large-scale structure building process increases. Gravity is able to enhance the anisotropies in the cloud and promote the formation of additional large-scale filaments, which eventually fragment to form stars \citep{vazquezsemadeni2007,2008MNRAS.389.1556B,2013IAUS..292....3M,vazquezsemadeni2019,2024arXiv240407301H}.

We find that reducing the virial parameter from $\als = 0.5$ to $\als =0.125$ increases the star formation rate per freefall time $\mr{SFR_{ff}}$ by a factor of $\sim2$. Reducing the virial parmeter further to $\als = 0.0625$ does not result in an additional increase in $\mr{SFR_{ff}}$, showing that $\als$ has a non-linear effect on $\mr{SFR_{ff}}$ (see Fig.~\ref{fig:sfr_bm}). Similar $\mr{SFR_{ff}}$ in simulations with $\als = 0.0625$ and $\als = 0.125$ imply that, at $\als \sim 0.1$, gravity has transformed most of the anisotropies or over-densities in the turbulent cloud into stars, i.e., the maximum possible efficiency of star formation at the given Mach number of $\mathcal{M} = 5$ is reached. Thus, lowering $\als$ further would not affect $\mr{SFR_{ff}}$ significantly. Nevertheless, the time at which star formation starts varies between the three models. The first stars form the latest in the simulations with $\als = 0.5$ and the earliest in the simulation with $\als = 0.0625$ (see Fig.~\ref{fig:sfr}). 

Previous numerical studies of how the initial turbulence influences the IMF are primarily centred on the aspect of the velocity dispersion, with most of the studies finding that the IMF is nearly insensitive to the velocity dispersion or Mach number (see \S\ref{sec:dis_imf}). However, analysing the effect of turbulence on the basis of the velocity dispersion alone is inadequate since changes in $\sigma_{\mr{v}}$ change both $\als$ and $\mathcal{M}$ at the same time, each of which has a unique influence on the cloud dynamics, structure, SFR, and IMF. Since in our simulations, we keep the velocity dispersion and therefore $\mathcal{M}$ fixed, our analysis is principally focused on the effect of changes in the virial parameter. Thus, our work complements previous studies on the influence of turbulence.

We find that the IMF has a weak dependence on the cloud virial parameter. On varying $\als$ in the simulations from $\als \sim 0.5$ to $\als \lesssim 0.1$, the peak mass of the IMF and the median mass decreases by $\sim 2$. We also find that the multiplicity fraction, $mf$, is higher for the lower cloud virial parameter cases, particularly in the sub-solar range. $mf$ in the sub-solar primary mass range for simulations with $\als = 0.0625$ and $0.125$ are higher than that for $\als = 0.5$ by a factor $\sim2$ (see Fig.~\ref{fig:Multiplicityfraction}). Such a difference is important given that this is the mass range where the peak of the IMF is expected. Nevertheless, all three models reproduce the trend of increasing multiplicity fraction with mass and the $mf$ estimates in individual mass intervals broadly agree with the values derived in observational surveys.   

We highlight that the non-linear dependence of the SFR and IMF on $\als$ needs to be considered (see Fig.~\ref{fig:sfr_bm} and Fig.~\ref{fig:avg_mass}). If $\als$ is investigated solely within a limited range of either very low or high values, its impact on the SFR and IMF will appear to be negligible, simply because either gravity or turbulence dominates over the other. To reach conclusive results, the effect of the virial parameter should be examined across a broad range of $\als$ values.

\section*{Acknowledgements}
We thank the anonymous referee for their quick and constructive review of the work. We would like to thank Blakesley Burkhart, Patrick Hennebelle, No$\acute{\text{e}}$ Brucy, and Mark Krumholz for discussions regarding SFR theory. S.S.M.~would like to acknowledge the useful discussions during `The Physics of Star Formation' winter school held in February 2024 at Les Houches, France. C.F.~acknowledges funding provided by the Australian Research Council (Discovery Project DP230102280), and the Australia-Germany Joint Research Cooperation Scheme (UA-DAAD). We further acknowledge high-performance computing resources provided by the Leibniz Rechenzentrum and the Gauss Centre for Supercomputing (grants~pr32lo, pr48pi and GCS Large-scale project~10391), the Australian National Computational Infrastructure (grant~ek9) and the Pawsey Supercomputing Centre (project~pawsey0810) in the framework of the National Computational Merit Allocation Scheme and the ANU Merit Allocation Scheme. The simulation software FLASH was in part developed by the DOE-supported Flash Center for Computational Science at the University of Chicago and the University of Rochester. This work makes use of the yt-project \citep{yt} and colormaps in the CMasher package \citep{cmasher}.

\section*{Data Availability}
The data used in this article is available upon reasonable request to the authors.



\bibliographystyle{mnras}
\bibliography{Bibliography} 




\appendix
\section{3D visualisation of the simulations}
\label{sec:app}
\begin{figure*}
    \centering
    \includegraphics[width=\textwidth]{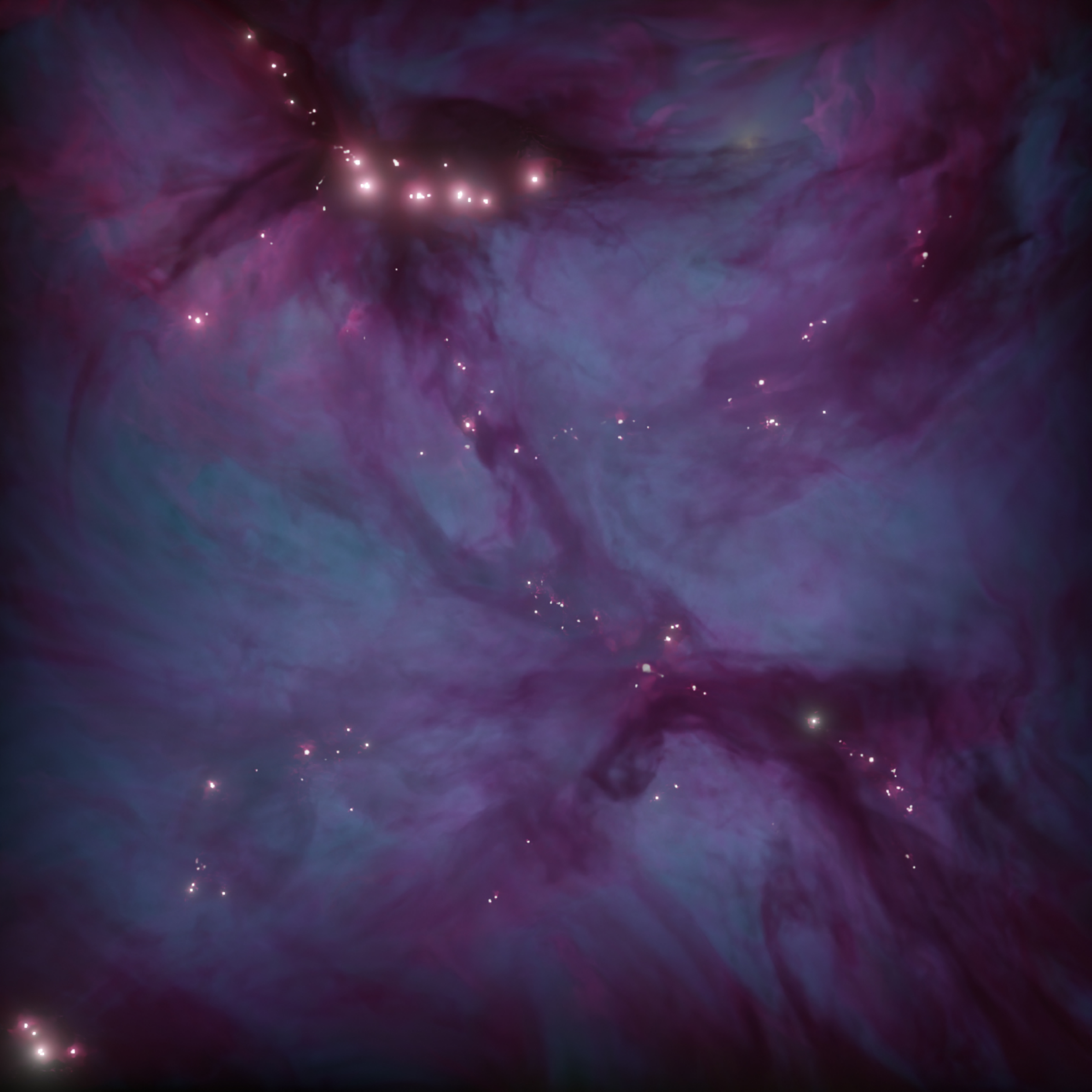}
    \caption{3D visualisation of the simulation with $\als = 0.0625$ at $\mathrm{SFE}=5\%$. The visualisation here presents the same simulation (same turbulence seed) as in the left panel of Fig.~\ref{fig:densmap}. Movies of the above visualisation can be seen at \href{https://sajaymathew.github.io/visualisations.html}{https://sajaymathew.github.io/visualisations.html}.} 
    \label{fig:3d_1}
\end{figure*}

\begin{figure*}
    \centering
    \includegraphics[width=\textwidth]{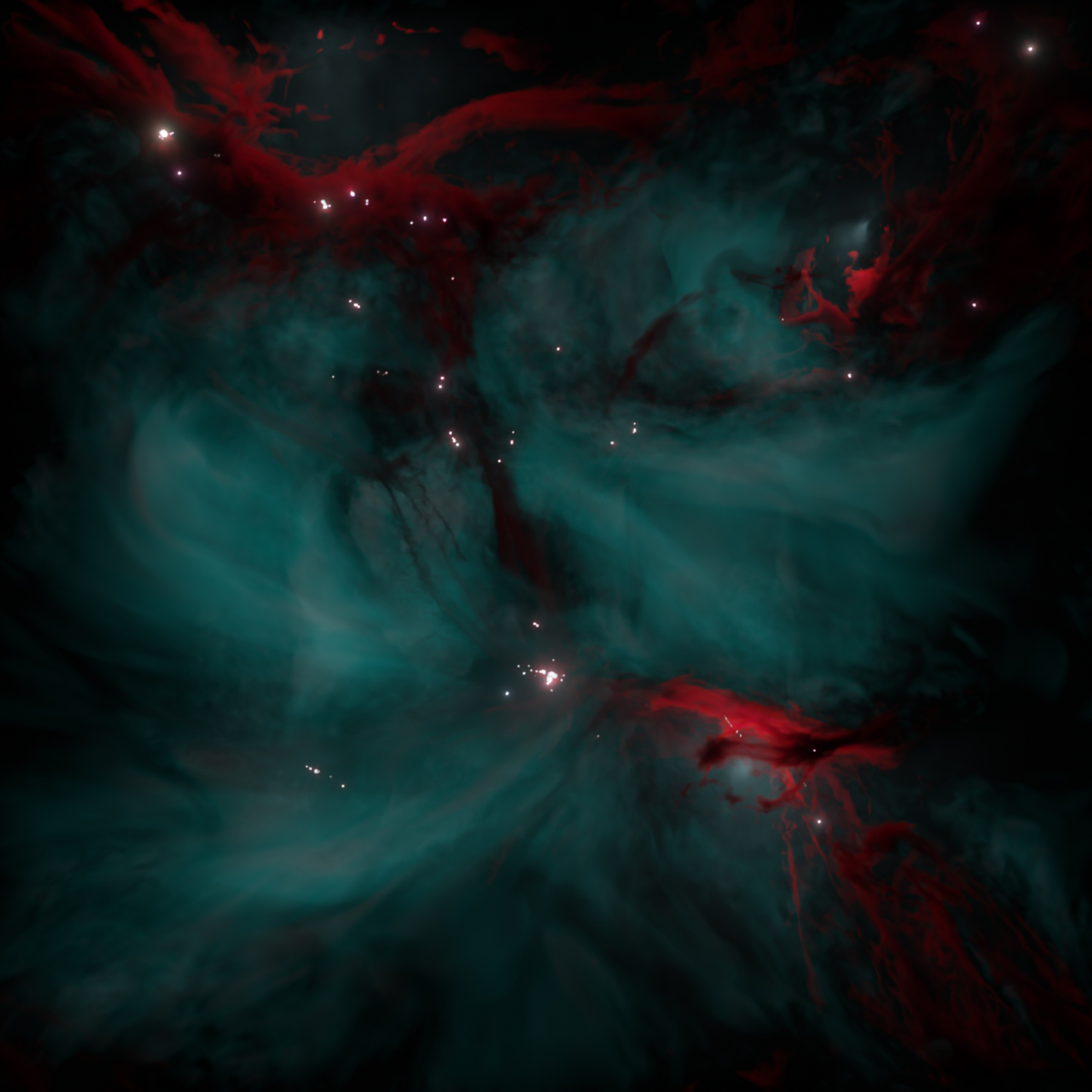}
    \caption{3D visualisation of the simulation with $\als = 0.125$ at $\mathrm{SFE}=5\%$. The visualisation depicts the same simulation as in the middle panel of Fig.~\ref{fig:densmap}.} 
    \label{fig:3d_2}
\end{figure*}

\begin{figure*}
    \centering
    \includegraphics[width=\textwidth]{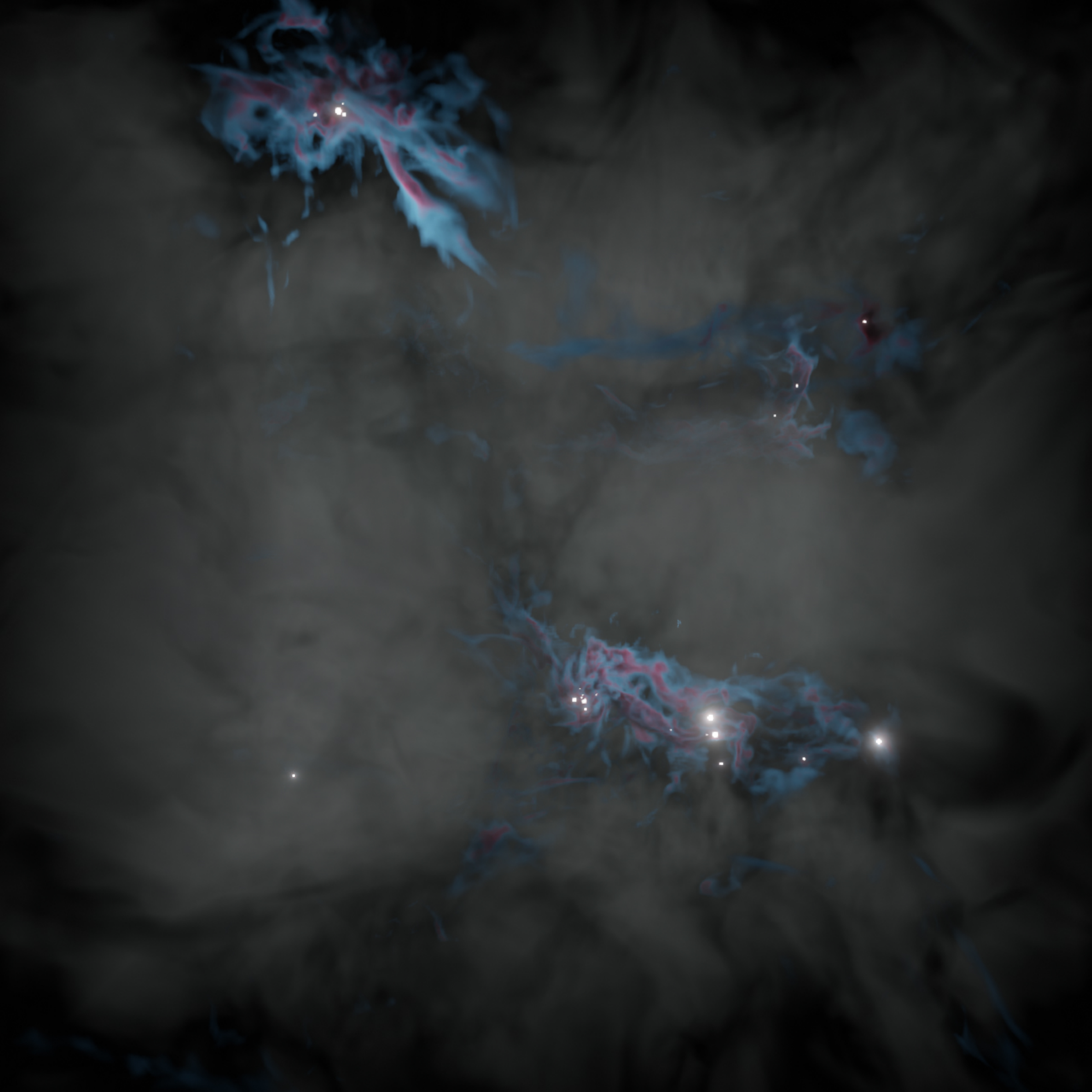}
    \caption{3D visualisation of the simulation with $\als = 0.5$ at $\mathrm{SFE}=5\%$. The visualisation shows the same simulation as in the right panel of Fig.~\ref{fig:densmap}.} 
    \label{fig:3d_3}
\end{figure*}

Fig.~\ref{fig:3d_1}, Fig.~\ref{fig:3d_2}, and Fig.~\ref{fig:3d_3} present 3D visualisations of the simulations shown in Fig.~\ref{fig:densmap}. Movies for these visualisations can be seen at \href{https://sajaymathew.github.io/visualisations.html}{https://sajaymathew.github.io/visualisations.html}.

\bsp	
\label{lastpage}
\end{document}